\documentclass[prd,aps,showpacs,twocolumn,a4paper,floatfix]{revtex4-1}
\newcommand{\hateq}{\,\,\hat{=}\,\,}
\newcommand{\re}{\mbox{Re}}

\newcommand{\Lie}{{\cal L}}

\newcommand\p{\partial}
\newcommand\mbeta{\mathring\beta}
\usepackage{mathrsfs}
\usepackage{graphicx,psfrag}
\usepackage{amsmath,amsfonts,amssymb}

\begin{document}

\title{Constraint preserving boundary conditions 
  for the Z4c formulation of general relativity}
\author{Milton Ruiz}
\email{milton.ruiz@uni-jena.de}
\author{David Hilditch}
\email{david.hilditch@uni-jena.de}
\author{Sebastiano Bernuzzi}
\email{sebastiano.bernuzzi@uni-jena.de}
\affiliation{Theoretical Physics Institute, University of 
Jena, 07743 Jena, Germany}


\begin{abstract}
We discuss high order absorbing constraint preserving boundary 
conditions for the Z4c formulation of general relativity coupled 
to the moving puncture family of gauges. We are primarily 
concerned with the constraint preservation and absorption properties 
of these conditions. In the frozen coefficient approximation, with 
an appropriate first order pseudo-differential reduction, we show 
that the constraint subsystem is boundary stable on a four 
dimensional compact manifold. We analyze the remainder of the 
initial boundary value problem for a spherical reduction of the 
Z4c formulation with a particular choice of the puncture gauge. 
Numerical evidence for the efficacy of the conditions is presented 
in spherical symmetry.
\end{abstract}


\maketitle
\tableofcontents

\section{Introduction}             
\label{section:Introduction}        

Numerical simulations of general relativity typically introduce 
an artificial time-like outer boundary. This boundary requires 
conditions which ought to render the initial boundary value 
problem (IBVP) well-posed. Well-posedness is the requirement 
that a unique solution of the IBVP exists and depends 
continuously upon given initial and boundary data.  

The most important approaches to demonstrate well-posedness of
the IBVP are the energy  and Laplace-Fourier transform methods
~\cite{Olsson05,Gustafsson95,Gundlach:2004jp,Nagy:2006pr}.
The energy method is  straightforward. In this approach one
constructs a  suitable norm for the solutions of the dynamical 
system. Using the equations of motion one can  estimate the 
growth of this norm in time. However, this technique in general 
cannot be used  if  the  system is not symmetric hyperbolic 
or if the boundary conditions are not maximally dissipative. 
Recently,  Kreiss {\it et al.} introduced in~\cite{Kreiss:2008ig}
a non-standard energy norm to prove that the IBVP for the 
second order systems of wave equations with
Sommerfeld-like boundary conditions is well-posed. 
The key idea in~\cite{Kreiss:2008ig} is to choose a particular
time-like direction in a way that the boundary conditions
are maximally dissipative ones. A different method is based on the frozen 
coefficient approximation. In this approach one freezes the coefficients 
of the equations of motion and the boundary operators. The IBVP is 
thus simplified to a linear, constant coefficient problem which 
can be solved using a Laplace-Fourier transformation. Sufficient 
conditions for the well-posedness of the frozen coefficient problem 
were developed by Kreiss in~\cite{Kreiss70} if the system is 
strictly hyperbolic. Using that theory, a smooth symmetrizer 
can be constructed with which well-posedness can be shown using 
an energy estimate in the frequency domain.  
Agranovich~\cite{Agranovich:1972} extended that theory to the 
case in which the system is strongly hyperbolic and the eigenvalues 
have  constant multiplicity. It is expected that, by using 
the theory of pseudo-differential operators~\cite{Taylor99b}, 
one can show well-posedness of the general problem.
In what follows we use the Laplace-Fourier approach to prove the
well-posedness of the IBVP for the constraint subsystem of Z4 with high 
order constraint preserving boundary conditions. 
The order of a boundary condition refers to the highest derivative 
of the metric or of the gauge variables contained therein.

Once the continuum boundary conditions (BCs)  are understood, one needs 
a strategy for their implementation in a numerical code. 
The numerical implementation is required to be stable and to converge
to the continuum solution at a certain rate.  If the system is symmetric
hyperbolic one can use, for instance, difference operators which satisfy
summation by parts schemes and penalty techniques to transfer information
through the outer boundary 
condition~\cite{Lehner:2005bz,carpenter94,carpenter99,carpenter02}.  
This allows the derivation of semi-discrete energy estimates which can 
guarantee the stability of the numerical implementation.
Nevertheless, in the general cases, even for a linear system, demonstrating 
that a numerical  approach to BCs will result in a stable scheme is 
difficult. In the absence of a proof of numerical stability one must 
rely on calculations for similar toy problems and on thorough 
numerical tests with simple data, in which problems can be identified 
locally at the boundary. Unfortunately naive discrete approaches are 
often numerically unstable. One issue is that a code usually requires 
more conditions than are given at the continuum level. 

The two most popular choices of formulation of general relativity (GR,
hereafter) in use in numerical relativity 
today are the generalized harmonic gauge (GHG) and the BSSN 
formulations~\cite{Friedrich85,Friedrich96,Shibata95,Baumgarte:1998te}. 
Significant progress has been made in the construction of both 
continuum and discrete BCs for the GHG formulation, see
{\em e.g.}~\cite{Szilagyi:2002kv,Lindblom:2005qh, Kreiss:2006mi,
Babiuc:2006wk,Motamed:2006uw,Rinne:2006vv,Ruiz:2007hg,Rinne:2008vn} 
and references therein.  For GHG the task is made relatively easy because 
the system has a very simple wave-equation structure in the principal part 
and furthermore because the constraints may be expressed as time 
derivatives of metric fields. The BSSN formulation is used to evolve 
both vacuum and matter space-times by a number of numerical relativity 
groups, see {\em e.g.}~\cite{Campanelli:2005dd,Baker:2005vv,Brugmann:2008zz,
Baiotti:2008ra,Tichy:2007hk,Etienne:2008re,Hinder:2008kv,Shibata:2009ad}
and references therein. This The system is taken in combination with the 
so-called moving puncture-gauge~\cite{Brugmann:2008zz}.
BCs for the BSSN formulation have received 
relatively little attention, although  recently, N\'u\~nez and 
Sarbach have proposed~\cite{Nunez:2009wn} constraint preserving 
boundary conditions (CPBCs) for this system. 
Recasting the dynamical system into a first order symmetric hyperbolic 
system, they are able to prove that the corresponding IBVP is
well-posed through a standard energy method, at least in the linearized case.
These boundary  conditions have not yet been implemented  numerically. 
Currently Sommerfeld BCs are the most common 
in use in applications, despite the fact that they are certainly not 
constraint preserving and it is not known whether or not they result in 
a well-posed IBVP. The problem is that the characteristic structure of 
puncture-gauge BSSN is more complicated than that of the GHG formulation, 
which makes the analysis difficult. Despite the fact that with Sommerfeld 
conditions the constraints do not properly converge, in applications they 
are robust and are currently not the dominant source of error in numerical 
simulations.

Another version of GR is the Z4 formulation~\cite{Bona:2003fj,
Bona:2003qn}. When coupled to the generalized harmonic gauge Z4 
is formally equivalent to GHG~\cite{Gundlach:2005eh}. Additionally 
it is possible to recover the BSSN formulation from Z4 by freezing 
one of the constraint variables. In this sense Z4 may be thought 
of as a generalization of both BSSN and GHG. Z4 has the advantage 
over GHG that it maintains sufficient gauge freedom that it may 
be coupled to the puncture-gauge, potentially allowing the 
evolution of puncture initial data as is standard with BSSN. To 
that end a conformal decomposition of the Z4 formulation (hereafter Z4c)
was recently presented~\cite{Bernuzzi:2009ex}. Unlike BSSN, the Z4 
formulation has a constraint subsystem in which every constraint 
propagates with the speed of light, which may be useful in 
avoiding constraint violation in numerical applications. It was 
shown that, at least in the context of spherical symmetry, 
numerical simulations performed with puncture-gauge Z4c have 
smaller errors than those performed with 
BSSN~\cite{Bernuzzi:2009ex}. However it was also found that 
Z4c is rather more sensitive than BSSN with regards to boundary 
conditions.  BCs compatible with the constraints
for a symmetric hyperbolic first order reduction of Z4 
were  specified and tested numerically in~\cite{Bona:2004ky,Bona:2010wn}.
Those conditions are of the maximally dissipative type and, 
therefore, the  well-posedness of the resulting IBVP was shown
by using a standard energy estimation. Nevertheless, Bona 
{\it et al.}  used in~\cite{Bona:2004ky,Bona:2010wn} 
harmonic slicing and normal coordinates to rewrite Z4 as a symmetric 
hyperbolic formulation. Therefore,  it is not clear if their  results
can be easily extended to the general case.

In this work we therefore specify BCs in
combination  with puncture-gauge Z4c and 
we show that the resulting IBVP is well-posed at least
in spherical scenarios. However,  since we are interested in
specifying CPBCs which can be used in 3D evolutions,
the well-posedness of IBVP for the constraint subsystem is
established in the general case. In addition, we study the effectiveness
of these conditions by performing numerical
evolutions in spherical symmetry.

We begin in section~\ref{section:Z4} with a summary of the 
Z4 formulation, and identify the BCs we would 
like to consider in our analysis. 
 We present the analytic setup
for our well-posedness analysis in section~\ref{section:Z4_setup}.
Section~\ref{section:proofs} contains our analytic results on 
BCs for Z4. We present our numerical results in
section~\ref{section:Numerical_Results}. 
We conclude  in section~\ref{section:Conclusion}. The principal 
ideas of the Kreiss theory are summarized in 
appendix~\ref{section:well-posed problems} and applied
to the wave equation with high order BCs
in appendix~\ref{section:Toy_model0}. We describe the  numerical implementation
    of the second-order CPBCs in Appendix~\ref{App:BC_imp}.

\section{The Z4c formulation}  
\label{section:Z4}            

In this section we present the Z4c formulation and the general
expressions for our BCs in detail. We also 
introduce notation for a $2+1$ decomposition in space, which 
we will use in the calculations in the following sections.

\subsection{Evolution equations and constraints}   

Following~\cite{Bernuzzi:2009ex}, in which the conformal Z4 
formulation was presented, we replace the Einstein equations 
with the expanded set of equations
\begin{eqnarray}
\p_t\gamma_{ij}&=& -\,2\,\alpha\,K_{ij} +\Lie_\beta\gamma_{ij} \,,
\label{eq:Z4_ADM_1}\\
\p_tK_{ij} &=&  -\,D_iD_j\alpha+\alpha\,
\left[R_{ij}-2\,K_{ik}\,K^k_j+K_{ij}\,K
\right.\nonumber\\
&&\left.+\,2\,\p_{(i}Z_{j)}\right]
+ 4\pi\,\alpha\,\left[\gamma_{ij}
(S-\rho)-2\,S_{ij}\right]
\nonumber\\
&&+\,\Lie_\beta K_{ij}\,,\\
\p_t\Theta &=& \alpha\,\left[\frac{1}{2}\,H + \p_kZ^k\right]
+\beta^i\,\Theta_{,i}\,,\label{eq:Theta_dot}\\
\p_tZ_{i}  &=& \alpha\,M_i + \alpha\,\Theta_{,i}+\beta^j\,Z_{i,j}\,,
\label{eq:Z4_ADM_2}
\end{eqnarray}
where $\Theta$ and $Z_i$ are constraints. The ADM equations are
recovered when the constraints $\Theta$ and $Z_i$ vanish.
The Hamiltonian and momentum constraints 
\begin{eqnarray}
H &=& R-K_{ij}\,K^{ij}+K^2 - 16\, \pi\,\rho = 0\,,\\
M^i &=& D_j\,\left(K^{ij}-\gamma^{ij}\,K\right) - 8\,\pi\, S^i=0\,,
\end{eqnarray}
evolve according to 
\begin{eqnarray}
\p_0H&\simeq& -2\,\p_iM^i\,,\label{eq:ham-Z4}\\
\p_0M_i&\simeq& -\frac{1}{2}\,\p_iH+\p^j\p_jZ_i-\p_i\p^jZ_j\,,
\label{eq:mom-Z4}
\end{eqnarray}
in the principal part, where we have used
\begin{eqnarray}
\p_0&=&\frac{1}{\alpha}\,\left(\p_t-\beta^i\p_i\right)\,.
\end{eqnarray}
Since we are concerned in this work only with the behavior of 
the BCs on the constraints, we have discarded 
the constraint damping scheme of~\cite{Gundlach:2005eh}.

\subsection{Conformal decomposition}   

In our numerical applications we evolve the Z4c system in the 
conformal variables $\chi,\tilde{\gamma}_{ij},\hat{K},
\tilde{A}_{ij}$ and $\tilde{\Gamma}^i$, defined by
\begin{eqnarray}
\tilde{\gamma}_{ij} = \gamma^{-\frac{1}{3}}\,\gamma_{ij}\,, 
&\qquad 
\hat{K} = \gamma^{ij}K_{ij}-2\,\Theta\,,\\
\chi = \gamma^{-\frac{1}{3}}\,,
&\qquad 
\tilde{A}_{ij}=\gamma^{-\frac{1}{3}}(K_{ij}-\frac{1}{3}
\gamma_{ij}K)\,,
\end{eqnarray}
and finally
\begin{eqnarray}
\tilde{\Gamma}^{i} &=&2\,\tilde{\gamma}^{ij}Z_j+
\tilde{\gamma}^{ij}\tilde{\gamma}^{kl}\tilde{\gamma}_{jk,l}\,.
\end{eqnarray}
The choice of conformal variables allows us to evolve 
puncture initial data whilst altering the underlying PDE  
properties of the formulation. In what follows we will 
use the shorthand
\begin{eqnarray}
\tilde{\Gamma}_{\textrm{\small{d}}}{}^i&=&
\tilde{\gamma}^{ij}\tilde{\gamma}^{kl}\tilde{\gamma}_{jk,l}=
\gamma^{\frac{1}{3}}\gamma^{ij}\gamma^{kl}\left(\gamma_{jk,l}
-\frac{1}{3}\gamma_{kl,j}\right)\,.
\end{eqnarray}
In terms of the conformal variables the evolution equations 
for the Z4c formulation become 
\begin{align}
\p_t \chi &= \frac{2}{3}\chi[\alpha(\hat{K}+2\Theta) - D_i\beta^i]\,,
\label{eq:Z4_decomp_first}\\
\p_t \tilde{\gamma}_{ij} &= -2\alpha\tilde{A}_{ij}+\beta^k
\tilde{\gamma}_{ij,k}+2\tilde{\gamma}_{(i|k}\beta^k_{,|j)}-\frac{2}{3}
\tilde{\gamma}_{ij}\beta^k{}_{,k}\,,\\
\p_t \hat{K}    &= -D^iD_i\alpha + \alpha[\tilde{A}_{ij}\tilde{A}^{ij}
+\frac{1}{3}(\hat{K}+2\Theta)^2]\nonumber\\
&+4\pi\alpha[S+\rho_{{\textrm {\tiny ADM}}}]+\beta^iK_{,i}\,,
\end{align}
the trace-free extrinsic curvature evolves with
\begin{align}
\p_t \tilde{A}_{ij} &= \chi[-D_iD_j\alpha
+\alpha (R_{ij}-8\pi S_{ij})]^{\textrm{tf}}\nonumber\\
& +\alpha[(\hat{K}+2\Theta)\tilde{A}_{ij} - 2\tilde{A}^k{}_i\tilde{A}_{kj}]
\nonumber\\
& +\beta^k\tilde{A}_{ij,k}+2\tilde{A}_{(i|k}\beta^{k}{}_{,|j)}
-\frac{2}{3}\tilde{A}_{ij}\beta^{k}{}_{,k}\,, 
\end{align}
and finally we have
\begin{align}
\p_t \tilde{\Gamma}^{i} &= -2\tilde{A}^{ij}\alpha_{,j}+2\alpha
[\tilde{\Gamma}^i_{jk}\tilde{A}^{jk}
-\frac{3}{2}\tilde{A}^{ij}\ln(\chi)_{,j}
\nonumber\\
&-\frac{1}{3}\tilde{\gamma}^{ij}(2\hat{K}+\Theta)_{,j}
-8\pi\tilde{\gamma}^{ij}S_j]
+\tilde{\gamma}^{jk}\beta^i_{,jk}\nonumber\\
&
+\frac{1}{3}\tilde{\gamma}
^{ij}\beta^k_{,kj}+\beta^j\tilde{\Gamma}^i_{,j}
-\tilde{\Gamma}_{\textrm{d}}{}^j\beta^i_{,j}+\frac{2}{3}
\tilde{\Gamma}_{\textrm{d}}{}^i\beta^j_{,j}\,.
\end{align}
The $\Theta$ variable evolves according to Eqn.~\eqref{eq:Theta_dot} with 
the appropriate substitutions 
Eqns.~(\ref{eq:Conf_Constr_1}-\ref{eq:Conf_Constr_2}). 
In the $\p_t\tilde{A}_{ij}$ equations we write  
\begin{align}
R_{ij} &= R^{\chi}{}_{ij} + \tilde{R}_{ij}\,,\\
\tilde{R}^{\chi}{}_{ij} &=
\frac{1}{2\chi}\tilde{D}_i\tilde{D}_j\chi+\frac{1}{2\chi}
\tilde{\gamma}_{ij}\tilde{D}^l\tilde{D}_l\chi\nonumber\\
&-\frac{1}{4\chi^2}\tilde{D}_i\chi\tilde{D}_j\chi-\frac{3}{4\chi^2}
\tilde{\gamma}_{ij}\tilde{D}^l\chi\tilde{D}_l\chi\,,\\
\tilde{R}_{ij} &=
 - \frac{1}{2}\tilde{\gamma}^{lm}
\tilde{\gamma}_{ij,lm} +\tilde{\gamma}_{k(i|}\tilde{\Gamma}
^k_{|,j)}+\tilde{\Gamma}_{\textrm{d}}{}^k
\tilde{\Gamma}_{(ij)k}+\nonumber\\
&\tilde{\gamma}^{lm}\left(2\tilde{\Gamma}^k_
{l(i}\tilde{\Gamma}_{j)km}+\tilde{\Gamma}^k_{im}
\tilde{\Gamma}_{klj}\right)\,.
\end{align}
The complete set of constraints are given by
\begin{align}
\Theta\,,  &  \qquad 
2 Z_i     = \tilde{\gamma}_{ij}\tilde{\Gamma}^{j}-
             \tilde{\gamma}^{jk}\tilde{\gamma}_{ij,k}\,,
\label{eq:Conf_Constr_1}\\
H       &= R -\tilde{A}^{ij}\tilde{A}_{ij}+\frac{2}{3}
           (\hat{K}+2\Theta)^2-16\pi\rho_{{\textrm {\tiny ADM}}}\,,\\
\tilde{M}^i &= \p_j\tilde{A}^{ij} + \tilde{\Gamma}^i{}_{jk}\tilde{A}^{jk}
             -\frac{2}{3}\tilde{\gamma}^{ij}\p_j(\hat{K}+2\Theta)\nonumber\\
            & -\frac{3}{2}\tilde{A}^{ij}(\log\chi)_{,j}\,,\\
D           &\equiv\ln(\det\tilde{\gamma})=0\,,\qquad
T            \equiv \tilde{\gamma}^{ij}\tilde{A}_{ij}=0\,.
\label{eq:Conf_Constr_2}
\end{align}
In our numerical evolutions the algebraic constraints $D$ and 
$T$ are imposed continuously during the numerical calculations.
It seems to improve the stability of the simulations significantly~\cite{Alcubierre08a}.

\subsection{Puncture gauge conditions}   

The most popular gauge choice in the numerical evolution of 
dynamical spacetimes is the puncture gauge. In introducing 
scalar functions ($\mu_L,\mu_S,\epsilon_\alpha,\epsilon_\chi$) 
the general form of the gauge (without introducing the additional 
field $B^i$) is
\begin{eqnarray}
\p_t\alpha&=&\beta^i\alpha_{,i}-\mu_L\,\alpha^2\,\hat{K}\,,
\label{eq:punc_alpha}\\
\p_t\beta^i&=&\beta^j\beta^i{}_{,j}+\mu_S\,\tilde{\Gamma}^i
-\eta\,\beta^i-\epsilon_\alpha\,\alpha\,\alpha^{,i}\nonumber\\
&+&\epsilon_\chi\,\tilde\gamma^{ij}\p_j\chi
\,.\label{eq:punc_beta}
\end{eqnarray}
Note that in this condition we have included a new term 
proportional to the spatial derivative of $\chi$. As we will 
show in Sec.~\ref{section:proofs}, the inclusion of that 
term allows us to prove the well-posedness of the IBVP for Z4 
in the frozen coefficient approximation. We refer to the choice
of shift $(\mu_S,\epsilon_\alpha,\epsilon_\chi)=(1,1,1/2)$ as 
the asymptotically harmonic shift condition because in preferred
coordinates in asymptotically flat spacetimes near infinity the 
condition asymptotes to the harmonic shift.

In evolutions of equal mass black holes the gauge damping 
parameter $\eta$ is usually taken as $2/M$, where $M$ is the 
ADM mass of the spacetime. Recently it has been shown that a 
spatially varying $\eta$ parameter may be helpful in the 
evolution of unequal mass binaries~\cite{Mueller:2010bu,Lousto:2010ut}.

\subsection{Fully second order form}   

The principal part of the Z4c formulation in fully second 
order form is given by
\begin{eqnarray}
\left(\p_0^2-\mu_L\p_i\p^i\right)\alpha&\simeq&0\,,
\label{eq:sec-lapse}\\
\left(\p_0^2-\gamma^{\frac{1}{3}}\frac{\mu_S}{\alpha^2}\p_j\p^j
\right)\beta^i
&\simeq&\left(\frac{\gamma^{\frac{1}{3}}\mu_S}{\alpha^2\mu_L}
-\epsilon_\alpha\right)\p_0\p^i\alpha\nonumber\\
&&\hspace{-1.5cm}
+\,\frac{1}{3\,\alpha}\,\left(
\frac{\gamma^{\frac{1}{3}}\mu_S}{2}-
\epsilon_\chi\right)\,\gamma^{jk}\p_0\p^i\gamma_{jk}\,,
\label{eq:sec-shift}\\
\left(\p_0^2-\p_l\p^l\right)\gamma_{ij}&\simeq&
\frac{1}{3}\gamma^{kl}\left(1-\frac{2\,\epsilon_\chi}{
\gamma^{\frac{1}{3}}\,\mu_S}\right)\p_{i}\p_{j}\gamma_{kl}\nonumber\\
&+&\frac{2}{\alpha}\left(
1-\frac{\alpha^2\epsilon_\alpha}{\gamma^{\frac{1}{3}}\mu_S}
\right)\p_i\p_j\alpha
\label{eq:sec-gamma}\\
&+&\frac{2}{\alpha}\left(
1-\frac{\alpha^2}{\gamma^{\frac{1}{3}}\mu_S}
\right)\gamma_{k(i}\p_{j)}\p_0\beta^k\,.
\nonumber
\end{eqnarray}
One may view the constraints $\Theta$ and $Z_i$ as being
defined by the gauge 
choice~(\ref{eq:punc_alpha}-\ref{eq:punc_beta}),
\begin{eqnarray} 
2\,\Theta &=&\frac{1}{\alpha\mu_L}\p_0\alpha
-\frac{1}{2}\gamma^{ij}\p_0\gamma_{ij}
+\frac{1}{\alpha}\p_i\beta^i\,,
\label{eq:def_theta}
\\
2\,Z_i&=&
\frac{1}{\mu_S\,\gamma^{1/3}}\,
\Big(\alpha\,\gamma_{ij}\,\p_0\beta^j
+\eta\,\beta_i+\epsilon_\alpha\,\alpha\,\alpha_{,i}\nonumber\\
&&\hspace{2cm}-\,\epsilon_\chi\,\gamma^{1/3}\,\p_i\chi
\Big)-{(\tilde\Gamma_\textrm{d})}_i\,.
\label{eq:def_Zi}
\end{eqnarray}
The principal parts of the constraint subsystem are just 
wave equations
\begin{eqnarray} 
\Box\Theta\simeq 0\,, &\quad& \Box Z_i \simeq 0\,,
\label{eq:sys-theta-Z}\\
\Box H\simeq 0\,,     &\quad& \Box M_i \simeq 0\,.
\label{eq:sys-ham-mom}
\end{eqnarray}
Following the approach of~\cite{Hilditch:2010wp} we will 
analyze the system starting in fully second order form. The equations 
of motion can be $2+1$ decomposed against the spatial unit 
vector $s^i$. We define the projection operator 
$q^i{}_j=\delta^i{}_j-s^is_j$. The equations of motion 
split into scalar, vector and tensor parts. The decomposed 
variables are written
\begin{eqnarray}
\gamma_{ss}&=&s^is^j\gamma_{ij}\,, 
\quad \gamma_{qq}=q^{ij}\gamma_{ij}\,,\\
\gamma_{sA}&=&s^iq^j{}_A\gamma_{ij}\,,\\
\gamma_{AB}^{\textrm{TF}}&=&\left(q^i{}_Aq^j{}_B
-\frac{1}{2}q_{AB}q^{ij}\right)\gamma_{ij}\,,\\
\beta_s&=&s^i\beta_i\,,\quad \beta_A=q^i{}_A\beta_i\,.
\end{eqnarray}
The metric and shift are reconstructed from the decomposed 
quantities by
\begin{eqnarray}
\gamma_{ij}&=&\big(q^A{}_{(i}q^B{}_{j)}-
\frac{1}{2}q^{AB}q_{ij}\big)\gamma^{\textrm{TF}}_{AB}
+q^A{}_{(i}s_{j)}\gamma_{sA}
\nonumber\\
&&s_is_j\gamma_{ss}+q_{ij}\gamma_{qq},\\
\beta_{i}&=&s_i\beta_s+q^{A}_{i}\beta_A.
\end{eqnarray}
Here and in what follows we use upper case Latin indices to 
denote projected quantities. 

\subsection{Characteristic variables}   

The standard parameters choice in the  gauge
conditions is $\mu_L=2/\alpha$, $\mu_S=3/4$ and 
$\epsilon_\alpha=\epsilon_\chi=0$. When Z4 is coupled to the 
puncture gauge it is typically strongly hyperbolic (necessary and 
sufficient for well-posedness of the initial value problem,
see {\it e.g.}~\cite{Alcubierre08a,Nagy:2004td}) 
except in a handful of special cases which for brevity we do 
not discuss here. The fully second order characteristic 
variables with $\epsilon_\alpha=\epsilon_\chi=0$ were presented 
in~\cite{Bernuzzi:2009ex}. Here we present them
 with the additional parameters. In the scalar sector
the characteristic variables are 
\small
\begin{align} 
&U^{\pm\sqrt{\mu_L}}=\p_0\alpha\pm\sqrt{\mu_L}\,\p_s\alpha\,,\\
&U^{\pm\lambda}=
\p_0\Lambda\pm\frac{\alpha^2\lambda}{\gamma^{1/3}\,\mu_S}\,\p_s\Lambda
\nonumber\\
&
-2\,\frac{\alpha^2-\gamma^{1/3}\,\mu_S}{(\lambda^2\,\alpha^2
-\gamma^{1/3}\,\mu_S)\,\alpha}
\left(\p_s\beta_s\pm\frac{\alpha^2\,\lambda}{\gamma^{1/3}
\,\mu_S}\,\p_0\beta_s\right)
\nonumber\\&
-
\frac{2\,(\lambda^2\,\alpha^2+\gamma^{1/3}\,\mu_L\,\mu_S)}{(\lambda^2-\mu_L)
(\lambda^2\,\alpha^2-\gamma^{1/3}\,\mu_L\,\mu_S)\,\alpha\,\mu_L}\,\left(\p_0\alpha
\pm\frac{\alpha^2\,\mu_L\lambda\,\epsilon_\alpha}{\gamma^{1/3}\,
\mu_S}\,\p_s\alpha\right)\nonumber\\
&
+\frac{2\,(\epsilon_\alpha-1)\,\alpha\,\lambda}
{(\lambda^2-\mu_L)(\lambda^2\,\alpha^2-\gamma^{1/3}\,\mu_S)}
\left(\p_s\alpha\mp\lambda\,\p_0\alpha\right)\nonumber\\
&+\frac{2\,(\lambda^2\,\gamma^{1/3}\,\mu_S+\alpha^2\,\epsilon_\alpha
\mu_L)}{(\lambda^2-\mu_L)(\lambda^2\,\alpha^2-\gamma^{1/3}\,\mu_S)
\,\alpha\,\mu_L}
\left(\p_0\alpha\pm\frac{\alpha^2\,\mu_L\lambda}{\gamma^{1/3}\,\mu_S}
\p_s\alpha\right)\,,\\
&U^{\pm 1}=\p_0\gamma_{qq}\pm\p_s\gamma_{qq}\,,\\
&U'^{\pm 1}=\p_0\beta_s\pm\frac{\gamma^{1/3}}{\alpha}\,\p_s\beta_s
+\epsilon_\alpha\,\p_s\alpha\pm \frac{\gamma^{1/3}\,\mu_S}
{\alpha^2\,\mu_L}\,\p_s\alpha\nonumber\\
&\mp\frac{\gamma^{1/3}\,\mu_S}{2\,\alpha}
\,\p_0\Lambda+\frac{\epsilon_\chi-2\,\gamma^{1/3}\,\mu_S}
{3\,\alpha}\,\p_s\Lambda\,.
\end{align}
\normalsize
with geometric speeds $\pm (\sqrt{\mu_L},\lambda,1,1)$ respectively, 
and where we have defined 
\begin{eqnarray}
\Lambda&=&\gamma_{ss}+\gamma_{qq}\,, \\
\lambda&=&\sqrt{\frac{2\,\left(2\,\gamma^{1/3}\,\mu_S
-\epsilon_\chi\right)}{3\,\alpha^2}}\,.
\end{eqnarray}
Note that if $\lambda$ vanishes the system is only weakly hyperbolic.
In the vector sector the characteristic variables are 
\begin{align}
U^{\pm\sqrt{\mu_S}}_{A}&=\p_0\beta_A
\pm\frac{\sqrt{\mu_S}\,\gamma^{1/3}}
{\alpha}\,\p_s\beta_A\,,\\
U^{\pm 1}_{A}&=\p_0\gamma_{sA}\pm\p_s\gamma_{sA}
-\frac{\alpha^2}{\gamma^{1/3}\mu_S}
(\p_0\beta_A\pm\p_s\beta_A)\,,
\end{align}
with speeds $\pm(\sqrt{\mu_S},1)$. Finally in the tensor sector 
we have simply
\begin{align}
U^{\pm 1}_{AB}&= \p_0\gamma_{AB}^{\textrm{TF}}
\pm\p_s\gamma_{AB}^{\textrm{TF}},
\end{align}
with speeds $\pm 1$.

Since the constraints $(\Theta,Z_i,H,M_i)$ satisfy wave equations 
in the principal part their characteristic variables are simply
\begin{align}
U^\pm_{\Theta}&=\p_0\Theta\pm\p_s\Theta\,,\quad
&U^\pm_{s}=\p_0Z_s\pm\p_sZ_s\,,\\
U^\pm_{A}&=\p_0Z_A\pm\p_sZ_A\,,\quad
&U^\pm_{H}=\p_0H\pm\p_sH\,,\\
U'^\pm_{s}&=\p_0M_s\pm\p_sM_s\,,\quad
&U'^\pm_{A}=\p_0M_A\pm\p_sM_A\,,
\end{align}
each with speeds $\pm 1$.

\subsection{High order absorbing constraint preserving 
boundary conditions}   

Following the notation of~\cite{Ruiz:2007hg} we investigate
the Z4c evolution equations on a manifold $M=[0,T]\times\Sigma$.
The three dimensional compact manifold $\Sigma$ has smooth 
boundary $\p\Sigma$. The boundary of the full manifold
$\mathcal{T}=[0,T]\times\p\Sigma$ is timelike and the three 
dimensional slices $\Sigma_t=\{t\}\times\Sigma$ are spacelike. 
The boundary of a spatial slice is denoted
$S_t=\{t\}\times\partial \Sigma$. 

We define a background metric ($\mathring{\alpha},
\mathring{\beta}_{i},\mathring{\gamma}_{ij}$),
\begin{eqnarray}
\textrm{d}\mathring{s}^2&=&
\mathring{g}_{ab}\,\textrm{d}x^a\,\textrm{d}x^b=
\nonumber\\
&-&\mathring{\alpha}^2\textrm{d}t^2
+\mathring{\gamma}_{ij}(\textrm{d}x^i+\mathring{\beta}^i\textrm{d}t)
(\textrm{d}x^j+\mathring{\beta}^j\textrm{d}t)\,.
\label{eq:backg-metric}
\end{eqnarray}
We assume the background 3-metric to be conformally flat for later convenience.
It can be written as
\begin{eqnarray}
\mathring{\gamma}_{ij}\,\textrm{d}x^i\,\textrm{d}x^j
&=&\mathring{\psi}^4\,\left(\textrm{d}r^2+r^2\textrm{d}\Omega^2\right)\,,
\end{eqnarray}
which defines the background isotropic radial coordinate $r$ and the 
metric on the two-sphere is $\textrm{d}\Omega^2$. We furthermore 
define $\mathring{n}^a$, the background future pointing unit normal 
to the slices $\Sigma_t$, and $\mathring{s}^a$, the unit background 
normal to the two-surface $\{t\}\times\p\Sigma$ as embedded in 
$\Sigma_t$. We are primarily concerned with absorbing conditions 
for the Z4c formulation as a PDE system. Constructing BCs explicitly 
related to the incoming gravitational radiation 
is left for future work. For simplicity our time-like and outgoing 
normal vectors~$(\mathring{n}^a,\mathring{s}^a)$ are therefore 
defined against the background metric. Therefore,
\begin{equation}
\mathring{g}_{ab}\,\mathring{n}^a\,\mathring{n}^b=-1\,,\quad 
\mathring{g}_{ab}\,\mathring{s}^a\,\mathring{s}^b=1\,,\quad 
\mathring{g}_{ab}\,\mathring{n}^a\,\mathring{s}^b=0\,.
\end{equation}

To finish formulating the BCs we define the background outgoing
characteristic vectors
\begin{eqnarray}
\mathring{l}^a&=& \frac{1}{\sqrt{2}}\,\left(\mathring{n}^a
+\mathring{s}^a\right)\,,\\
\mathring{k}^a&=& \frac{1}{\sqrt{2}}\,\left(\mathring{n}^a
+\sqrt{\nu_s}\,\mathring{s}^a\right)\,,
\label{eq:nullvector-k}\\
\mathring{j}^a&=& \frac{1}{\sqrt{2}}\,\left(\mathring{n}^a
+\sqrt{\nu_T}\,\mathring{s}^a\right)\,,\\
\mathring{m}^a&=& \frac{1}{\sqrt{2}}\,\left(\mathring{n}^a
+\sqrt{\mu_L}\,\mathring{s}^a\right)\,,
\label{eq:nullvector-m}
\end{eqnarray}
where $\nu_s$ and $\nu_T$ are the characteristic speeds 
associated with equation~(\ref{eq:sec-shift}) in the scalar and 
vector sector, respectively. Since the constraints $\Theta,Z_i$ satisfy 
wave equations, constraint preserving, absorbing BCs 
in the linear regime around the background are given by 
\begin{equation}
\label{eq:general_CPBCs}
\boxed{
\left(r^2\,\mathring{l}^a\p_a\right)^{L}\Theta \hateq 0\,,\quad 
\left(r^2\,\mathring{l}^a\p_a\right)^{L} Z_i \hateq 0\,,}
\end{equation}
where $L\ge 0$ is an integer and $\hateq$ denotes equality in 
the boundary $\mathcal{T}$.  Note that the above conditions
can be considered a generalization those recently proposed by 
Bona {\it et.al.} in~\cite{Bona:2004ky,Bona:2010wn} which correspond 
to $L=0$ (see also~\cite{Bayliss:1980,Buchman:2006xf}).

We assume that both the physical and background metrics are 
sufficiently close to flat so that the full system has ten incoming 
characteristic variables at the boundary, which determines the number 
of BCs we may specify. The boundary 
conditions~(\ref{eq:general_CPBCs}) give four of the total. We 
take for the remainder 
\begin{eqnarray}
\left(r^2\,\mathring{m}^a\p_a\right)^{L+1}\alpha&\hateq h_\alpha\,,
\label{eq:BCs-alpha}\\
\left(r^2\,\mathring{k}^a\p_a\right)^{L+1}\beta_s&\hateq h_s\,,
\label{eq:general_BCs_gauge_first}\\
\left(r^2\,\mathring{j}^a\p_a\right)^{L+1}\beta_A&\hateq h_A\,,\\
\left(r^2\,\mathring{l}^a\p_a\right)^{L+1}\gamma_{AB}^{\textrm{TF}}&
\hateq h_{AB}^{\textrm{TF}}\,,
\label{eq:BCs_lastII}
\end{eqnarray}
where $h_\alpha,h_i\,,h^{\textrm{TF}}_{AB}$ are given boundary data. 
Since the above conditions are not tailored to the characteristic 
structure of the puncture gauge Z4c system there is no guarantee 
that they will absorb all outgoing fields. We leave detailed 
examination of the absorption properties, and possible important
modifications, of the 
conditions~(\ref{eq:BCs-alpha}-\ref{eq:BCs_lastII}) to future 
work and focus in the following discussion on well-posedness of 
the constraint subsector. In the following sections 
conditions~(\ref{eq:BCs-alpha}-\ref{eq:BCs_lastII}) are 
considered only in a spherical reduction of the system. 
The constraint absorption properties of 
conditions~\eqref{eq:general_CPBCs} are studied in our numerical 
tests.

\section{Analytical setup}
\label{section:Z4_setup}                     

In this section we discuss the strategy adopted to prove
well-posedness, namely the frozen coefficient approximation. 
This simplification is  necessary since one can show that
our system is not symmetric hyperbolic and our BCs
are not maximally dissipative. We also outline 
the cascade method which can be used for general proofs, 
provided that the equations of motion of the system have a 
special structure.

\subsection{Frozen coefficient approximation}   

Once the BCs are specified, one should
determine whether or not the resulting IBVP 
from~(\ref{eq:sec-lapse}-\ref{eq:sec-gamma}) 
with~(\ref{eq:general_CPBCs}-\ref{eq:BCs_lastII})
is well-posed.  This can be done by using the frozen coefficient 
technique, where one considers a high-frequency perturbation  
of a smooth background solution. This regime is the relevant 
limit for analyzing the continuous dependence of the solution 
on the initial data. By considering such a perturbation one can 
detect  the appearance of high frequency modes with exponential 
growth. Therefore, if the IBVP is well-posed in the frozen 
coefficient approximation, it is expected that the problem is 
well-posed in the non-linear case. In this limit, the  
coefficients of the equations of motion and the boundary operators 
can be frozen to their value at an arbitrary point. So, the 
problem is simplified to a linear, constant coefficient problem 
on the half-space which can be solved explicitly by using a 
Fourier-Laplace transformation~\cite{Gustafsson95,kreiss2001}.  
This method yields a simple algebraic condition (see 
appendix~\ref{section:well-posed problems}) which is necessary 
for the well-posedness of the IBVP.

Following~\cite{Ruiz:2007hg}, we perform a coordinate 
transformation which leaves $\Sigma_t$ invariant, such that one 
can rewrite the background metric~(\ref{eq:backg-metric}) at 
the point $p$ in the form
\begin{eqnarray}
\left.\textrm{d}\mathring{s}^2\right|_p&=&-\,\textrm{d}t^2
+\left(\,\mathring\beta\,\textrm{d}t+\textrm{d}x\right)^2
+\textrm{d}y^2+\textrm{d}z^2\,,
\label{eq:backg-metricP}
\end{eqnarray}
where $\mathring{\beta}$ corresponds to the normal component of 
the shift vector at $p$. According with this approximation, one can
assume  that the BC is a plane. Therefore, the
spatial manifold  becomes  $\Sigma=\{(x,y,z)\in \mathbb{R}^3:x>0\}$.
We denote the flat spatial metric at 
$p$ by $\eta_{ij}$.  Regarding the above metric, the time-like and
outgoing normal vectors~$(\mathring{n}^a,\mathring{s}^a)$ are 
\begin{equation}
\mathring{n}^a\p_a=\p_t-\mathring\beta\,\p_x\,,\qquad
\mathring{s}^a\p_a=-\mathring\beta\,\p_x\,.
\end{equation}
Besides, by using~(\ref{eq:backg-metricP}) and 
choosing 1+log slicing $\mu_L=2/\alpha$ and fixing $\mu_S=1$
in the shift condition one can rewrite the equations of 
motion~(\ref{eq:sec-lapse}-\ref{eq:sec-gamma}) 
in the frozen coefficient approximation at a point $p$ in the 
form
\begin{align}
\left(\p^2_0-2\,\p^l\p_l\right)\alpha&=0\,,
\label{eq:full-sec-z4-alpha}\\
\left(\p_0^2-\p^l\p_l\right)\beta^i&=
\left(\frac{1}{2}-\epsilon_\alpha\right)\p^i\p_0\alpha
\nonumber\\
&+\frac{1}{3}\left(\frac{1}{2}-\epsilon_\chi\right)
\eta^{jk}\p_0\p^i\gamma_{jk}\,,
\label{eq:full-sec-z4-beta}\\
\left(\p_0^2-\p^l\p_l\right)\gamma_{ij}&=\frac{1}{3}
(1-2\,\epsilon_\chi)\eta^{kl}\p_i\p_j\gamma_{kl}\nonumber\\
&+2(1-\epsilon_\alpha)\p_i\p_j\alpha\,.
\label{eq:full-sec-z4-metric}
\end{align}
Note that with the additional choice of asymptotically harmonic
shift $(\epsilon_\alpha,\epsilon_\chi)=(1,1/2)$ the resulting IBVP 
for the above system with boundary 
conditions~(\ref{eq:general_CPBCs}-\ref{eq:BCs_lastII})
has a cascade property; the gauge sector $(\alpha,\beta_i)$ is
coupled with the metric only through the BCs.
One can analyze  the resulting IBVP for the gauge sector and
then use it as a source in the remaining system. Nevertheless,
with the standard choices~$(\epsilon_\alpha,\epsilon_\chi)=(0,0)$,
all the variables are coupled to each other in the bulk 
as well as at the boundary. In this case, one should analyze 
the full system simultaneously. We will present the analytical  
results for arbitrary value of these parameters in~\cite{Hilditch:2010prep}.

\subsection{2+1 decomposition}     
\label{sec:2+1}                                         

To rewrite the above system with 
BCs~(\ref{eq:general_CPBCs}-\ref{eq:BCs_lastII})
as a set of cascade of wave problems, we perform
a $2+1$ decomposition against the spatial unit
vector $s^i=-\hat{e}_x$. Thus, the lapse
satisfies 
\begin{eqnarray}
&&\hspace{-0.8cm}
\left[\p_t^2-2\,\mathring\beta\,\p_t\p_x-(2-\mathring\beta^2)\,\p_x^2
-2\,\p^A\p_A\right]\alpha=0\,,
\label{eq:2+1Z4_a}\\
&&\hspace{-0.8cm}
\left[\p_t-\left(\sqrt{2}+\mathring{\beta}
\right)\,\p_x\right]^{L+1} 
\alpha\hateq h_\alpha\,.
\label{eq:2+1Z4_ac}
\end{eqnarray}
The wave problem for $\beta_{s}$ and $\beta_{A}$ is obtained 
once we project the system
\begin{eqnarray}
&&\hspace{-1.0cm}
\Big[\p_t^2-2\,\mathring\beta\,\p_t\p_x-(1-\mathring\beta^2)\p_x^2
-\p^C\p_C\Big]\beta_{i}=\nonumber\\
&&\hspace{4.5cm}-\frac{1}{2}\,\p_0\p_i\alpha\,,
\label{eq:Z4_betas_2+1}\\
&&\hspace{-1.0cm}
\left[\p_t-\left(1+\mathring{\beta}\right)
\,\p_x\right]^{L+1}\beta_{i}\hateq h_{i}\,,
\label{eq:Z4_betas_2+1BC}
\end{eqnarray}
along $s^i$ or along the transverse directions, respectively.
Note that the lapse is only coupled with the  equation~(\ref{eq:Z4_betas_2+1})
in the bulk. One could naively think in to analyze the 
lapse subsystem independently and use it as a source 
in~(\ref{eq:Z4_betas_2+1}). Therefore, the resulting
global estimate for the gauge sector will contain more derivatives of the 
lapse than of the shift. In general, it can spoil the estimate when one considers 
lower order terms  which appear in the nonlinear case.
In order to prevent it one should consider the wave problems for the lapse
and the shift simultaneously.

The system for the $\gamma^{\textrm{TF}}_{AB}$  is
\begin{eqnarray}
&&\hspace{-0.8cm}
\Big[\p_t^2-2\,\mathring\beta\,\p_t\p_x-(1-\mathring\beta^2)\p_x^2
-\p^C\p_C\Big]\gamma^{\textrm{TF}}_{AB}=0\,,
\label{eq:Z4_gammaAB_2+1}\\
&&\hspace{-0.8cm}
\left[\p_t-\left(1+\mathring{\beta}\right)\,
\p_x\right]^{L+1} \gamma^{\textrm{TF}}_{AB}
\hateq h^{\textrm{TF}}_{AB}\,.
\label{eq:Z4_gammaAB_2+1BC}
\end{eqnarray}
According to the results presented in the 
appendix~\ref{section:Toy_model0}, it is straightforward to prove  
this problem  is well-posed. Since the above subsystem
is decoupled completely from the rest of the system, it can be 
considered  as a given function in the remaining wave problems.
On the other hand, by introducing  the trace variable $\Lambda$ and considering the 
equation~(\ref{eq:def_theta}) as a definition of the constraint 
$\Theta$, we obtain
\begin{eqnarray}
&&\hspace{-1.0cm}
\left[\p_t^2-2\,\mathring\beta\,\p_t\p_x-\left(1-
\mathring\beta^2\right)\p_x^2
-\p^A\p_A\right]\Lambda=0\,,
\label{eq:EOM-Lambda}\\
&&\left[\p_t-\left(1+\mathring{\beta}\right)\,\p_x\right]^{L}
\Big[(\p_t-\mathring{\beta}\,\p_x)
(\alpha-\Lambda)-\nonumber\\&&\hspace{2.5cm}
2\,\p_x\beta_s+2\,\p^A\beta_A\Big]
\hateq0\,.
\label{eq:BC-Lambda}
\end{eqnarray}
Since this system is also decoupled from the rest of the metric sector,
one can analyze the resulting IBVP and after that, use it as a given
source for the other problems. Finally, the remaining equation of motions 
are again obtained through the projection of the wave equation
\begin{eqnarray}
&&\hspace{-0.5cm}
\Big[\p_t^2-2\,\mathring\beta\,\p_t\p_x-(1-\mathring\beta^2)\,\p_x^2
-\p^C\p_C\Big]\gamma_{is} =0\,,
\label{eq:eomgammasA}
\end{eqnarray}
along $s^i$ or along the transverse directions respectively.
By virtue of Eqn.~(\ref{eq:def_Zi}) the BC for 
$\gamma_{ss}$ is
\begin{eqnarray}
&&\hspace{-0.8cm}
\left[\p_t-\left(1+\mathring{\beta}\right)\p_x\right]^L\Big[(\p_t-
\mathring{\beta}\,\p_x)\beta_s
-\p^A\gamma_{sA}-\nonumber\\&&\hspace{2.5cm}
\p_x\left(\alpha-\gamma_{ss}+\Lambda/6\right)
\Big]\hateq 0\,,
\label{eq:Z4_gammasbc_2+1}
\end{eqnarray}
and finally, the BC for $\gamma_{sA}$ is
\begin{eqnarray}
&&\hspace{-1cm}
\left[\p_t-\left(1+\mathring{\beta}\right)\,\p_x\right]^L
\Big[(\p_t-\mathring{\beta}\,\p_x)\beta_A-
\p^B\gamma^{\textrm{TF}}_{AB}+
\nonumber\\&&\hspace{1cm}
\p_x\gamma_{sA}+\p_A\left(\alpha-\Lambda/3+\gamma_{ss}/2\right)
\Big]\hateq0\,.
\label{eq:Z4_gammaAbc_2+1}
\end{eqnarray}
This subsector does not have the cascade property. The
equation of motion for $\gamma_{ss}$ and $\gamma_{sA}$ 
are decoupled, but their BCs are coupled to each 
other. Therefore, one should consider the mutually coupled wave 
problems~(\ref{eq:eomgammasA}-\ref{eq:Z4_gammaAbc_2+1}) 
simultaneously.

In the following section, we consider the case with 
trivial initial data. Notice that this is not a real 
restriction.  One can always  treat the case of general 
initial data by considering, for instance,  the transformation 
$\bar{u}(t,x) = u(t,x) -g(t)\,f(x)$, where $g(t)$ is a smooth function 
with compact support such that $g(0)=1$. Therefore, $\bar{u}(t,x)$ 
satisfies the same  IBVP as $u(t,x)$  with modified sources and 
trivial initial data~\cite{Gustafsson95}. 

\section{Well-posedness Results}
\label{section:proofs}   

This section contains our analysis of well-posedness 
for different BCs. As we have mentioned before,
we consider the IBVP for the constraint subsystem on the 
manifold $M=[0,T]\times\Sigma$ and we just analyze the 
corresponding IBVP for the dynamical Z4c variables
on a spherical scenario. To do this, we explicitly 
solve the boundary problem using the Laplace-Fourier transformation. 
Kreiss presented in~\cite{Kreiss70} sufficient conditions 
for the well-posedness  in the frozen coefficient 
approximation. The key result in~\cite{Kreiss70} is the 
construction of a smooth symmetrizer for the problem for 
which well-posedness can be shown using an energy estimate 
in the frequency domain. We summarize the principal ideas 
of the Kreiss theory in the 
appendix~\ref{section:well-posed problems}.

\subsection{Constraint subsystem}    
\label{sec:CPBCs-constriant}                                         

Consider the IBVP for $\Theta$ with first order BCs (L=0)
and,  for a moment, let us assume  inhomogeneous  BCs, 
{\it i.e.} $\Theta\hateq q$, where $q$ is a given boundary data.
One can show that the equation of 
motion~(\ref{eq:sys-theta-Z}) and the boundary
for this variable can be rewritten in the 
form
\begin{displaymath}
\left[(s^2+\omega^2)-2\,\mathring\beta\,s\,\p_x-(1-\mathring\beta^2)\,\p_x^2
\right]\tilde\Theta=0\,,\qquad
\tilde\Theta\hateq \tilde q\,,
\label{eq:theta-eq-LF}
\end{displaymath}
where $\tilde\Theta$ and $\tilde q$ denote the Laplace-Fourier transformation of $\Theta$ and $q$ 
with respect to the directions $t$ and $x^A$ respectively and 
$\omega=\sqrt{\omega^2_y+\omega_z^2}$.  Following~\cite{Ruiz:2007hg,Kreiss:2006mi}
we rewrite the above system in the form~(\ref{eq:generalform-U}-\ref{eq:generalform-U-BC})
by  introducing  the variable
\begin{eqnarray}
D\tilde\Theta= \frac{1}{\kappa}\,\left(\p_x\tilde\Theta+
\gamma^2_\mu\,\mathring\beta\,s\,\tilde\Theta\right)\,,
\end{eqnarray}
with $\tilde{W} = \left(\tilde \Theta,D\tilde{\Theta}\right)^T$,
$L(s,\omega)=(1,0)$ and
\begin{eqnarray}
M(s,\omega) &=& \kappa\,\left( \begin{array}{cc}
-\gamma^2\,\mathring\beta\,s'  & 1 \\
\gamma^4\,\lambda^2
 & -\gamma^2\,\mathring\beta\,s'
\end{array} \right)\,,
\label{eq:MLF-Mat}
\end{eqnarray}
where we have defined $\gamma=1/\sqrt{1-\mathring\beta^2}$,  $\kappa=\sqrt{|s|^2+\omega^2}$,
$s'=s/\kappa$, $\omega'=\omega/\kappa$ and $\lambda^2=s'^2
+\gamma^{-2}\,\omega'^2$. 
The corresponding eigenvalues and eigenvectors of $M=M(s,\omega)$ are
\begin{equation}
\tau_\pm=-\kappa\,\gamma^2\,(s'\,\mbeta\mp\lambda)\,,\qquad
\hat{e}_\pm=(1,\pm\gamma^2\,\lambda)^T\,.
\end{equation}
Using this, it can be shown that the  $L_2$ solution of the system(\ref{eq:generalform-U-BC}) is 
given by
\begin{equation}
  \tilde{W}(s,x,\omega) = \sigma\,\hat e^-\,\text{exp}(\tau^-\, x)\,
\label{eq:general_sol_theta}
\end{equation}
where~$\sigma$ is a complex integration constant which 
is determined by introducing~(\ref{eq:general_sol_theta}) into the 
boundary, {\it i.e.} this constant satisfies $\sigma=\tilde q$.
Therefore, it follows immediately that
\begin{equation}
|\tilde W(s,0,\omega)|\leq C\,|\tilde q|\,,
\label{eq:fcpbcTheta}
\end{equation}
where $C$ is a strictly positive constant. Provided that the eigenvectors in
the solution~(\ref{eq:general_sol_theta}) are normalized in such a way that they
remain finite as $\omega\rightarrow 0$, as $\omega \rightarrow \pm\infty$ and
as $|s|\rightarrow \infty$ then we conclude that the resulting
IBVP for~(\ref{eq:sys-theta-Z}) with the first order boundary
condition~(\ref{eq:general_CPBCs}) is well-posed for trivial
initial data. By inverting the Laplace transformation and using the Parseval's
identity, we obtain an $L_2$ estimate of the form
\begin{equation}
\int_0^T\|W(\cdot,t)\|^2_\Sigma\,dt\leq C_T\,\int_0^T\|q\|^2_{\partial\Sigma}\,dt\,,
\label{eq:esTheta}
\end{equation}
in the interval $0\leq t\leq T$ for some strictly positive constant $C_T>0$.
Note that if the constraint $\Theta$ is satisfied initially and we consider
trivial boundary data $q=0$, the above inequality implies that the constraint
is satisfied everywhere and at each time. An equivalent estimate for $Z_i$ holds.

We generalize our previous analysis to consider BCs 
which depend on second or higher order derivatives of the constraints.
Recently, it has been  shown  that those conditions 
reduce the amount of  spurious reflections at the 
boundary~\cite{Buchman:2007pj,Rinne:2008vn}.
In fact, the BCs~(\ref{eq:general_CPBCs}) are Dirichlet 
conditions for the constraint subsystem~(\ref{eq:sys-theta-Z}), which 
means that the constraint violations that leave the bulk  are reflected 
at the boundary.  Therefore, Let us consider the wave problem for the $\Theta$ with 
high order BCs ($L\geq 1$).
Note that according to the appendix~\ref{section:Toy_model0}, it is possible
to rewrite the boundary matrix $L=L(s,\omega)$ in the form
\begin{equation}
L(s,\omega)=\frac{1}{2}\,\left(a_+^{L}+a_-^{L}, -
\,\frac{a_+^{L}-a_-^{L}}{\lambda\,\gamma^{2}}\right)\,,
\label{eq:hoTheta-bc}
\end{equation}
where $a_{\pm}=s'\pm\lambda$.
The $L_2$ solution of this system is given by~(\ref{eq:general_sol_theta})
but now the integration constant $\sigma$ satisfies
\begin{equation}
a_+^L\,\sigma=\tilde q\,.
\end{equation}
According to appendix~\ref{section:Toy_model0}, one can show that there is 
a strictly positive constant $\delta>0$ such that
$|a_+|^L=|s'+\lambda|>\delta$. Therefore, it follows that the system satisfies
the  estimate~(\ref{eq:fcpbcTheta}). We conclude that the solution
of the system (\ref{eq:sys-theta-Z}) with high order constraint preserving
BCs~(\ref{eq:general_CPBCs}) is well-posed and it
satisfies~(\ref{eq:esTheta}). The IBVP for $Z_i$ with these BCs
can be treated in a similar manner.

\subsection{Gauge subsystem} 
\label{sec:CPBCsGauge}       

Consider now the spherical reduction of the wave
problems~(\ref{eq:2+1Z4_a}-\ref{eq:Z4_betas_2+1BC})
for the lapse and the shift with first order BCs, {\it i.e.}
conditions~(\ref{eq:2+1Z4_ac}) and (\ref{eq:Z4_betas_2+1BC})
with $L=0$. By introducing the first order reduction variables
\begin{eqnarray}
D\tilde\alpha&=&\frac{1}{\kappa}\,\left(\p_x+\gamma^2_\alpha\,
\mathring{\beta}\,s\right)\tilde\alpha\,,\\
D\tilde\beta_s&=&\frac{1}{\kappa}\,\left(\p_x+\gamma^2\,
\mathring{\beta}\,s\right)\tilde\beta_s\,,
\end{eqnarray}
it can be possible to rewrite those wave problems in the 
form~(\ref{eq:generalform-U-BC}) with
\begin{widetext}
\begin{eqnarray}
\tilde{W} &=& 
\left( \begin{array}{c}
\tilde\alpha\\ 
D\tilde\alpha\\ 
\tilde\beta_s\\ 
D\tilde\beta_s\\
\end{array} \right)\,,
\qquad
M(s)= \kappa\,\left( \begin{array}{cccc}
-\gamma_\alpha^2\,\mathring\beta\,s'    &  1  & 0 & 0 \\
\gamma^4_\alpha\,s'^2_\alpha    &  -\gamma^2_\alpha\,\mathring\beta\,s'  & 0&0 \\
0&0&-\gamma^2\,\mathring\beta\,s'    &  1\\
-2\,s'^2\,\mathring\beta\,\gamma^2\,\gamma^4_\alpha&s'\,(2+\mathring\beta^2)
\,\gamma^2\,\gamma^2_\alpha/2&\gamma^4\,s'^2    &  -\gamma^2\,\mathring\beta\,s'\\
\end{array} \right)\,,\nonumber\\
L(s)&=&\left( \begin{array}{cccc}
\sqrt{2}\,s'&-\gamma^{-2}_\alpha &0 &0\\
0&0&s'&-\gamma^{-2}\\
\end{array} \right)\,,\qquad
\tilde{g}\, =
\left( \begin{array}{c}
\tilde g_\alpha\\
\tilde g_{s}\\
\end{array} \right)=
\frac{1}{\kappa}
\,\left(\begin{array}{c}
(\sqrt{2}-\mathring\beta)\,
\tilde h_{\alpha}\\
(1-\mathring\beta)
\,\tilde h_{s}
\end{array} \right)\,.\nonumber
\end{eqnarray}
\end{widetext}
Here we have defined $s'=s/\kappa$, $\kappa=|s|$ and
\begin{eqnarray}
\gamma^{-2}_\alpha&=&2-\mathring\beta^2\,,\qquad
\gamma^{-2}=1-\mathring\beta^2\,.
\end{eqnarray}
The $L_2$ solution of the above system is given by
\begin{eqnarray}
\tilde{W}(s,x,\omega) &=&\sum_{i=1}^2
\sigma_i\,\hat e_i^-\,\textrm{exp}\left(\tau_i^-\,x\right)\,,
\label{eq:genralsol-gauge}
\end{eqnarray}
where $\sigma_i$ are the complex integration constants and
$\tau^-_i$ and $e^-_i$ the negative eigenvalues of $M$ and $e_i^-$ the 
corresponding eigenvectors.
By replacing the solution into the boundary condition, it is  
possible to show that  $\sigma_i$ satisfy
\begin{align}
\sigma_\alpha&=\frac{\tilde g_\alpha}{2\,\sqrt{2}\,s'}\\
\sigma_s&=\frac{\tilde g_s}{2\,s'}-
\frac{\tilde g_\alpha\,\left[(1-\mbeta)\,(2+\sqrt{2}+\mbeta\,(1+\sqrt{2})\right]
\,\gamma^2_\alpha}{8\,s'}\,.
\end{align}
By the same argument as above, it is straightforward by using the 
triangle inequality that the system is boundary stable and therefore 
an equivalent estimate as~(\ref{eq:esTheta}) holds for the gauge subsystem
with first order BCs.

Next, let us consider high order BCs for this subsystem. 
To give an estimate for the solution
of the gauge subsystem with high order conditions
we rewrite them in an algebraic form.
Therefore, by using the procedure presented  in appendix~\ref{section:Toy_model0}
and by virtue of the equations of
motion~(\ref{eq:full-sec-z4-alpha}-\ref{eq:full-sec-z4-beta}),
one can rewrite the boundaries~(\ref{eq:Z4_betas_2+1}-\ref{eq:Z4_betas_2+1BC})
with $L=0$ in the form  $\mathcal{L}\tilde{W} = A\,\tilde W$ where
\begin{displaymath}
A=\left( \begin{array}{cccc}
\sqrt{2}\,s'   & -\gamma_\alpha^{-2}  & 0 & 0 \\
-2\,s'^2\gamma_\alpha^2    & \sqrt{2}\,s' & 0 & 0 \\
0 &  0& s' & \gamma^{-2} \\
2\,s'^2\,\mbeta\,\gamma^4_\alpha\, &  s'(2+\mbeta^2)\,\gamma^2_\alpha/2& -\gamma^2\,s^2&s' \\
\end{array} \right)\,.
\end{displaymath}
We have defined the boundary operator by 
$\mathcal{L}=(\mathcal{L}_{\tiny{\sqrt{2}}},\mathcal{L}_{\tiny{1}})^T$
with
\begin{equation}
\mathcal{L}_\mu=(\mu-\mbeta)\,s'-\frac{1}{\kappa\,\gamma^2_\mu}\,\p_x\,.
\end{equation}
By iteration, explicitly one obtains $\mathcal{L}^{L+1}\tilde
W=A^{L+1}\,\tilde W$. Here we have defined
\begin{widetext}
\begin{displaymath}
A^{L+1}=\frac{1}{2}\,\left( \begin{array}{cccc}
a^{L+1}_+&-a^{L+1}_+/(\sqrt{2}\,s'\,\gamma_\alpha^2)&0&0\\
-\sqrt{2}\,s'\,\gamma^2_\alpha\,a^{L+1}_+&a^{L+1}_+&0&0\\
-F(\mbeta)\,\gamma^4_\alpha\,a^{L+1}_+/\gamma^2
&G(\mbeta)\,\gamma^2_\alpha\,a^{L+1}/(s'\,\gamma^2)&b^{L+1}_+&-b^{L+1}_+/(s'\,\gamma^2)\\
H(\mbeta)\,s'\,\gamma^4_\alpha\,a^{L+1}_+&-J(\mbeta)\,\gamma^2_\alpha\,a^{L+1}_+& -s'\,\gamma^2\,b^{L+1}_+&b^{L+1}_+\\
\end{array} \right)\,,
\end{displaymath}
 \end{widetext}
where $a_+=2\,\sqrt{2}\,s'$ and $b_+=2\,s'$ are the eigenvalues of
the matrix $A=A(s')$ and  $F(\mbeta),\cdots,J(\mbeta)$ are certain shorthands 
for combinations of $\mbeta$ only. Thus, the high order BCs
for the gauge subsystem  can be rewritten in the
form~(\ref{eq:generalform-U-BC})  with the boundary matrix given by
 \begin{widetext}
\begin{equation}
L(s,\omega)=
\frac{1}{2}\,\left( \begin{array}{cccc}
a^{L+1}_+&-a^{L+1}_+/(\sqrt{2}\,s'\,\gamma_\alpha^2)&0&0\\
-F(\mbeta)\,\gamma^4_\alpha\,a^{L+1}_+/\gamma^2
&G(\mbeta)\,\gamma^2_\alpha\,a^{L+1}/(s'\,\gamma^2)&b^{L+1}_+&-b^{L+1}_+/(s'\,\gamma^2)
\end{array} \right)\,,
\end{equation}
 \end{widetext}
and the given data
\begin{equation}
\qquad
\left( \begin{array}{c}
\tilde g_\alpha\\
\tilde g_{s}\\
\end{array} \right)=
\frac{1}{\kappa^{L+1}}
\,\left(\begin{array}{c}
(\sqrt{2}-\mathring\beta)^{L+1}\,
\tilde h_{\alpha}\\
(1-\mathring\beta)^{L+1}
\,\tilde h_{s}
\end{array} \right)\,.
\end{equation}
The $L_2$ solution of the system is given
by~(\ref{eq:genralsol-gauge}). Nevertheless, the complex integration 
constants now satisfy
\begin{eqnarray}
\sigma_\alpha&=&\frac{\tilde g_\alpha}{2\,a^{L+1}}\,,\\
\sigma_s&=&\tilde g_\alpha\,\left[
\frac{F(\mbeta)+\sqrt{2}\,G(\mbeta)\,\gamma_a^4}{4\,b^{L+1}\,\gamma^2}-\frac{2\,(1+\sqrt{2})\,(1-\mbeta)}{8\,(2-\sqrt{2}\mbeta)\,a^{L+1}}\right]+
\nonumber\\&&
\frac{\tilde g_s}{2\,b^{L+1}}\,.
\end{eqnarray}
Since $\re(s')>0$, $a^{L+1}$ and $b^{L+1}$ are proportional to $s'$ and the
remain coefficient are constants, it follows that the system is boundary
stable and therefore well-posed. Similar arguments apply to the wave problem
of the metric components.

\section{Numerical applications}
\label{section:Numerical_Results}

The necessity of CPBCs for the Z4c system is not only motivated 
by the fundamental requirement of having a mathematically 
well-posed system, but also by the numerical evidence of artifacts 
related to the implementation of inadequate BCs. 
The property of full propagation of the constraints is both  
a strength and a weakness of the Z4 evolution system. On one hand,
by a comparison with BSSN it was shown in~\cite{Bernuzzi:2009ex} 
that this property reduces constraint violations on the grid.
On the other it makes the BCs a more important 
issue, because if the numerical boundary condition introduces 
large constraint violations, perhaps as spurious reflections,
then the violation may propagate inside the domain and swamp the
numerical solution.

\begin{figure}[t]
  \begin{center}
    \includegraphics[width=0.5\textwidth]{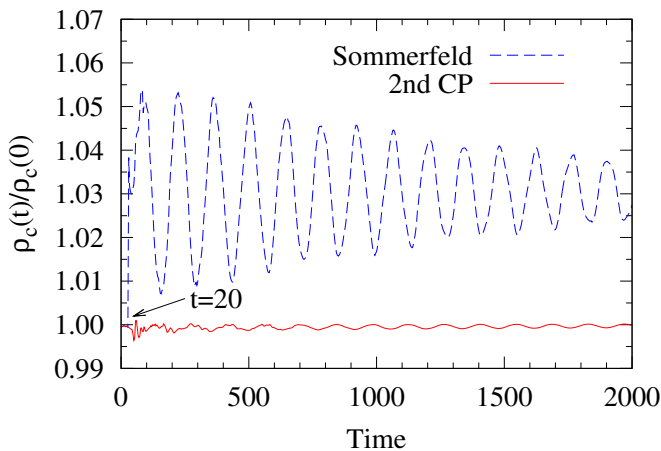}\\
    \caption{ \label{fig:rhoc} Radial oscillations of the 
      central rest-mass density of a compact star in time. 
      When Sommerfeld conditions are used a perturbation from 
      the boundary (at $r_{\rm out}=20$) hits the center of the 
      star (at $t=20$) and further perturbs it.
} 
  \end{center}
\end{figure}

As an example of numerical artifact, pointed out but only 
briefly discussed in~\cite{Bernuzzi:2009ex}, we show in 
Fig.~\ref{fig:rhoc} the time evolution of the central 
rest-mass density of a equilibrium model of spherical compact 
star obtained with the Z4c formulation (coupled to general 
relativistic hydrodynamics equations) and two different 
BCs: Sommerfeld and CPBCs
\footnote{In~\cite{Bernuzzi:2009ex} maximally dissipative 
conditions were used to cure the problem, CPBCs are the natural 
extension.}.
The central density is expected to remain constant in time but 
the truncation error of the numerical scheme causes small 
oscillations around the initial value that converge away with 
resolution. This is clearly visible when CPBCs are used. The 
frequency of these oscillations corresponds to the proper radial 
mode of the star. When Sommerfeld conditions are used however 
the constraint violation at the boundary propagates into the 
domain and perturbs the star as soon as it becomes causally 
connected with the outer boundary. The perturbation from the 
boundary alters the mean value of the central density. The 
``boundary'' perturbation does not converge (or converges much 
slower than the interior, see Sec.~\ref{subsec:flat}) so becomes 
the dominant error when the grid is refined. At later times the 
oscillations are damped by the hydrodynamical interaction between 
the fluid and artificial vacuum, but the mean value of the 
central density is forever modified from the initial one. Such 
boundary artifacts can thus dramatically move the solution far 
away from the initial configuration in phase space, despite the 
fact that at late times the constraint violation is very small. 
Preliminary 3D simulations with the Jena \textsc{BAM} 
code~\cite{Brugmann:2008zz} of single neutron stars, single 
puncture and binary black holes showed features very similar to 
the spherical results and in some cases instabilities triggered by 
the boundary.

In the following sections we discuss numerical results in 
spherical symmetry focusing only on the boundary 
conditions~(\ref{eq:general_CPBCs}). Since the practical implementation 
of BCs in a code is also an issue, in Sec.~\ref{sec:implementation} 
we summarize the method we used as well as other standard approaches. 
We use the code of~\cite{Bernuzzi:2009ex}. For more information on 
our numerical method, spherical reduction of the equations please 
refer to appendix A of that reference. We perform several tests, in 
each case with Sommerfeld and constraint preserving conditions. To 
examine stability and the effect of the BCs we consider 
simulations with a very close outer boundary 
($r_{\rm out}\simeq 20\,M$) and compare the results with a 
\emph{reference simulation}~\cite{Rinne:2007ui,Ruiz:2007hg}, in which the 
outer boundary is placed far away ($r'_{\rm out}\simeq 1000\,M$) 
from the origin. Since the boundary of the reference solution 
is causally disconnected on the time scale of the simulations 
with closer boundary, the BCs in the reference run have no 
importance. Results are presented for moderate resolution, about 
$\Delta r\sim 0.12$; but higher resolution runs as well as very 
long-term (hundreds of thousands of crossing times) simulations  
were performed showing the same behavior and no numerical 
instabilities. To monitor the global constraint violation we 
define the quantity:
\begin{equation}
  C \equiv \sqrt{H^2 + M^i M_i + \Theta^2 + Z^i Z_i} \ ,
\label{eq:Cmonitor}
\end{equation}
and we will refer to it as the \emph{constraint monitor}.
We will often make use of 2-norms of quantities:
\begin{equation}
  ||C(\cdot,t)||_2 \equiv \sqrt{\int dr\, r^2 C(r,t)^2} \ ,
\label{eq:norm2}
\end{equation}
where in practical computations the integral is performed on 
the grid by the trapezium rule. For a fair comparison with 
the ``near-boundary'' solution, the norm of the reference 
solution is taken only on  the domain, $r\in(0,r_{\rm out})$.

Since most of the  analytical results were obtained with 
the new asymptotically harmonic shift condition, we examine 
this gauge as well as the standard puncture gauge. In all 
cases we have found comparable results (see {\it e.g.} 
Fig.~\ref{fig:flat_gauge}) and therefore we will focus primarily  
on the standard puncture gauge. We therefore aim to give at 
least some numerical evidence of well-posedness in those cases 
where we are unable to demonstrate strong mathematical results.

A brief description of the tests performed and their aim follows
below. 

\paragraph*{Perturbed flat spacetime.} Evolution of constraint 
violating initial data on flat space. Here we focus on 
convergence and constraint absorption. We find near-perfect 
constraint transmission of the constraints when using the 
second order CPBCs. We devote the most attention to this test
because the effect of the BCs are clearest
in the absence of other sources of error. 

\paragraph*{Star spacetime.} Evolution of a stable compact 
star. In the Sommerfeld case, non convergent reflections from 
the boundary effect the dynamics of the star. The absorbing 
CPBCs completely solve this problem.

\paragraph*{Black hole spacetime.} Evolution of black hole 
initial data. The robustness and performance of CPBCs are 
tested against black hole spacetimes with different 
initial data and gauges. In particular we evolve a single 
puncture and Schwarzschild with a Kerr-Schild slicing. 

We use geometrical units $c=G=1$ everywhere, in case of matter 
spacetime dimensionless units are adopted setting $M_\odot=1$, 
in case of black hole spacetime $M_{\rm bh}=1$, while in case 
of flat spacetime a mass scale remains arbitrary.

\subsection{Numerical implementation of boundary conditions}   
\label{sec:implementation}

The literature contains many suggestions for the 
implementation of BCs, of which we highlight 
a small subset here. 

\paragraph*{Populate ghostzones.} One example is the recipe 
of~\cite{Calabrese:2005fp}, under which numerical stability 
of the shifted wave equation was proven. The idea is to write 
the BCs on the first order in time second 
order in space characteristic variables (which only implicitly 
contain time derivatives) and populate ghostzones so that the 
desired continuum boundary condition is satisfied. Ghostzones 
not determined by the BCs are simply populated 
by extrapolation. Since this method relies in an essential way 
on altering spatial derivatives locally, it is not clear how 
to apply the approach if one is computing spatial derivatives 
pseudo-spectrally. Furthermore the recipe may not give a unique 
prescription when one is given a system of equations.

\paragraph*{Summation by parts.}  As we have mentioned before,
with the summation by parts  schemes and penalty techniques
a quantity that mirrors the continuum energy for a symmetric 
hyperbolic system is constructed on the discrete system. Since we 
do not rely on an energy (symmetric hyperbolicity) method in our 
well-posedness analysis we are not able to construct a summation 
by parts finite difference scheme that guarantees stability. 
Analysis and applications in numerical relativity can be found 
in~\cite{Calabrese:2003vy,Calabrese:2003vx,Lehner:2005bz,
Diener:2005tn,Seiler:2008hm,Taylor:2010ki}.

\paragraph*{BCs as time derivatives of 
evolved fields.} The remaining approach we consider assumes 
that one may take spatial derivatives everywhere in a spatial 
slice. Inside a pseudo-spectral method spatial derivatives are 
naturally defined everywhere. On the other hand when 
approximating derivatives by finite differences, one may either 
use extrapolation to populate ghostzones, or take lop/one-sided 
differences near the boundary. The two methods are equivalent. 
To express the BCs on time derivatives of the 
metric one starts by substituting the definitions of the 
time-reduction variables (for example $\Theta, Z_i$ and $K_{ij}$) 
into the BCs. Then, if higher than first order 
time derivatives are still required, one may define a set of 
auxiliary reduction variables. One may use the equations of motion 
of the auxiliary variables in combination with the boundary 
conditions to eliminate spatial derivatives of auxiliary variables 
so that they may be confined to the 
boundary~\cite{Bayliss:1980,Novak:2002hn}. This approach was  
tested in the Caltech-Cornell SpEC code~\cite{Rinne:2008vn}, but it is
currently not used for binary black hole simulations. 
Here we perform evolutions only up to second order, and so do not 
need to use auxiliary variables in the boundary. 

Our numerical tests are performed with a spherical reduction
of the Z4c system. The numerical implementation of the boundary 
conditions in spherical symmetry is described in 
appendix~\ref{App:BC_imp}.

\subsection{Perturbed flat spacetime}
\label{subsec:flat}

\begin{figure}[t]
  \begin{center}
    \includegraphics[width=0.5\textwidth]{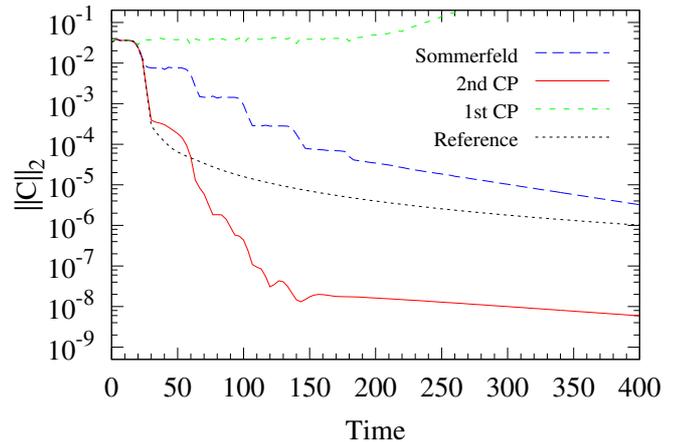}\\
    \caption{ \label{fig:flat_C} Constraint violation in flat
      spacetime test. The 2-norm of the constraint monitor is showed
      in time for different BCs implemented. The same quantity for the
      reference simulation is showed.}
  \end{center}
\end{figure}

In this test an initial constraint violating Gaussian 
perturbation is prescribed in variable $\chi$. During the 
evolution it propagates, reaches the boundary and if completely 
absorbed, the system relaxes to the Minkowski solution. This 
does not happen in practice because reflections are always 
present. Here we investigate the magnitude of these reflections 
and compare CPBCs of 1st and 2nd order with standard Sommerfeld 
conditions.

Figure~\ref{fig:flat_C} shows the 2-norm of the constraint 
monitor in time, results from the reference simulation are 
also reported. All the data agree up to around $t=20$, {\it i.e.} 
when the solution is traveling through the grid. After that time 
the following happens: (i) the constraint violation remains 
almost constant for 1st order CPBCs (green dotted line) and 
eventually causes the simulation to crash; (ii) the constraint 
violation decreases for Sommerfeld (blue dashed line) initially 
not monotonically (notice the four plateaus) then, after $t=180$, 
reaching a monotonic behavior; (iii) the constraint violation for 2nd 
order CPBCs (continuous red line) is smaller than  
Sommerfeld, plateaus are less clear but a monotonic behavior 
is also reached after $t=180$; (iv) the reference solutions 
(black dotted line) agree among themselves (remember that on the 
plotted timescale the boundary of the reference solution is 
disconnected from the spatial domain under consideration) and 
decrease monotonically but in absolute value less than 2nd order
CPBCs.

The interpretation of these observation is quite obvious, at 
least for points (i)-(iii): 1st order CPBCs simply 
\underline{do not} absorb the perturbation, which is 
entirely reflected back; Sommerfeld BCs are affected by partial 
reflections and at each crossing time a smaller portion of the 
wave comes back into the domain until an almost complete 
absorption; 2nd order CPBCs immediately absorb the largest 
portion of the outgoing wave. These statements are quantified 
below. 

Let us briefly discuss point (iv). A close look at the evolution of 
the constraint monitor in space shows a systematic drift from zero 
due to a back-scattering effect of the perturbation leaving the grid.
This is responsible for the larger values of the constraint violation. 
Such an effect can not be simulated in the closer domain case 
because the CPBCs cut completely the incoming modes of the solution, 
and, effectively, two different numerical spacetimes are simulated 
in the two cases.

An important point is to quantify the absorption properties of the
BCs under consideration. To this end we consider the characteristic
fields associated to the $\Theta$ given by:
\begin{equation}
U^\pm_\Theta = \partial_0 \Theta \pm \Theta_{,s} \ .
\label{eq:theta _modes}
\end{equation}
They can be regarded as incoming and outcoming modes of the 
solution.
Thus we define an experimental \emph{reflection coefficient} 
(inspired by \cite{Buchman:2006xf,Buchman:2007pj}) defined as the ratio of the Fourier 
modes of the characteristic fields at the boundary: 
\begin{equation}
R \equiv \frac{|\tilde{U}^-_\Theta(k)|}{|\tilde{U}^+_\Theta(k)|}. 
\label{eq:refcoef}
\end{equation}

\begin{figure}[t]
  \begin{center}
    \includegraphics[width=0.5\textwidth]{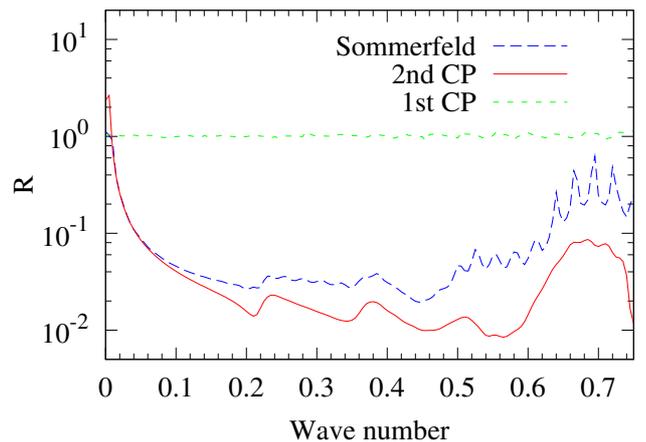}\\
    \caption{ \label{fig:flat_R} Experimental reflection coefficient
      in flat spacetime test. The experimental reflection coefficient, 
      defined in Eq.~(\ref{eq:refcoef}), is plotted versus the wave 
      number for different BCs implemented.}
  \end{center}
\end{figure}

Figure~\ref{fig:flat_R} shows that $R\sim 1$ for 1st order CPBCs, while
the behavior of 2nd order CPBCs is qualitatively similar to Sommerfeld.
Since they do not absorb the constraint violation, in what follows, we 
discard the 1st order CPBCs. 

\begin{figure}[t]
  \begin{center}
    \includegraphics[width=0.5\textwidth]{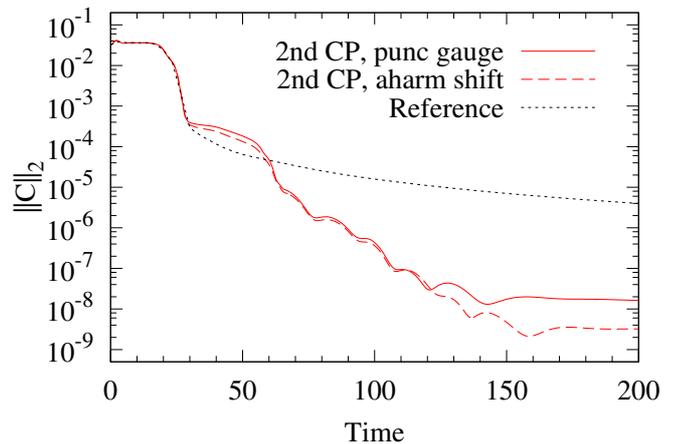}\\
    \caption{ \label{fig:flat_gauge} Constraint violation in flat
      spacetime test. Comparison of 2nd order CPBCs
      with puncture and asymptotically harmonic shift.}
  \end{center}
\end{figure}

The results presented so far refer to the puncture gauge. As stated 
at the beginning of the section, basically no significant differences 
are found when the asymptotically harmonic shift is employed. 
Figure~\ref{fig:flat_gauge} shows clearly this fact, we do not further 
comment on the asymptotically harmonic gauge until Sec.~\ref{subsec:bh}.

\begin{figure}[t]
  \begin{center}
    \includegraphics[width=0.5\textwidth]{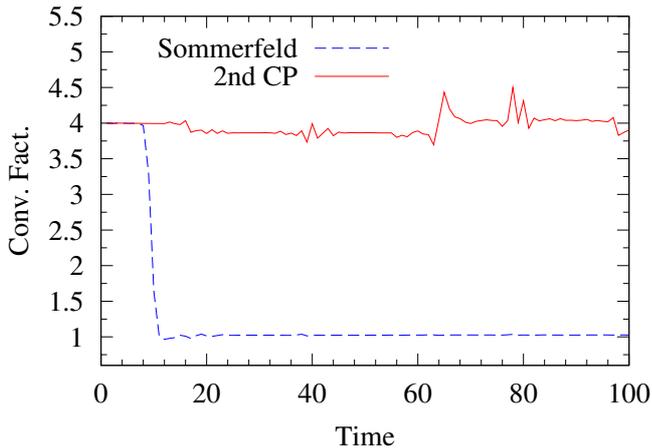}\\
    \caption{ \label{fig:flat_conv} Convergence factor in flat
      spacetime test. The self-convergence factor is compute from the
      2-norms of three simulations at different resolutions 
      and showed in time for 2nd order CPBCs and
      Sommerfeld BCs.}
  \end{center}
\end{figure}

Finally, we present convergence results. In figure~\ref{fig:flat_conv} 
the experimental self-convergence factor is plotted in time. While the 
physical solution is traveling on grid, $t<20$, the scheme is fourth 
order convergent as expected. Afterwards the numerical solution consists 
of the boundary reflections only and, while in case of Sommerfeld 
reflections are first order accurate, in case of CPBCs they maintain 
fourth order convergence up to $t\sim100$. For later times (not shown 
in the plot) the absolute value of the solution (and of the reflections) 
is so small that only noise is seen. We remark that the order of 
extrapolation used to fill the ghosts points and the finite difference 
operators in the two approaches are the same, so the differences are 
really due only to BCs. In order to obtain these convergence results, 
a non-staggered grid must be used because the staggered grid 
converges to the continuum domain at only first order in the grid 
spacing. In the following applications we use staggered grids, since 
this is commonly done in 3D codes.

\subsection{Star spacetime}
\label{subsec:star}

In this test we evolve stable spherical star initial data
(see~\cite{Bernuzzi:2009ex} for detail). At the beginning of 
the section we already discussed one of the main drawback of 
the use of Sommerfeld BCs with Z4c. As it is evident from 
Figure~\ref{fig:rhoc}, the use of CPBCs is not optional but 
absolutely necessary to obtain reliable results. In a more 
complicated/dynamical scenario in fact such artifacts could
be hidden or erroneously interpreted as real physics.

In this paper we are presenting results obtained without the
Z4 constraint damping scheme~\cite{Gundlach:2005eh}. One may 
however consider using the constraint damping scheme with a 
sufficiently large computational domain to suppress 
perturbations from the boundary. In our experience 
this is not only a inefficient cure (especially 
in 3D simulations) but an ineffective one. Our simulations 
indicate that the required damping coefficients are quite 
large, possibly because the perturbation from the boundary is 
typically not a high-frequency perturbation, and the damping 
scheme is most effective on high frequency perturbations.
Constraint damping is analytically understood in the linear 
regime and high frequency approximation, in a more general 
situation the indiscriminate use of constraint damping may 
lead to undesirable effects ({\it e.g.} qualitatively similar 
to large artificial dissipation).

\begin{figure}[t]
  \begin{center}
    \includegraphics[width=0.5\textwidth]{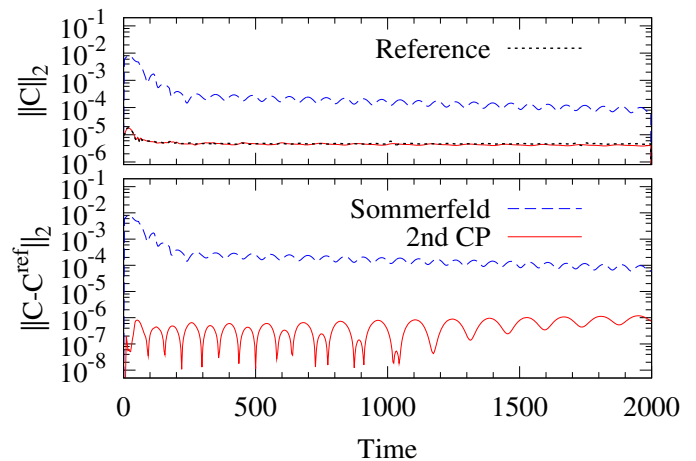}\\
    \caption{ \label{fig:star_C} Constraint violation in star 
      spacetime test. Upper panel: The 2-norm of the constraint 
      monitor is plotted in time for different BCs implemented. 
      The same quantity for the reference simulation is shown 
      (black dotted line). 
      Bottom panel: The 2-distance of the constraint monitor with 
      the reference simulation is showed in time for different 
      BCs implemented.}
  \end{center}
\end{figure}

In the upper panel of Figure~\ref{fig:star_C} we show the 2-norm 
of the constraint monitor for Sommerfeld, 2nd order CPBCs and the 
reference solution. It is evident that CPBCs are closer to the 
reference solution. In the bottom panel the distance from the 
reference solution is plotted showing that CPBCs lead to an 
improvement of approximately 2 to 4 orders of magnitude.

\subsection{Black hole spacetime}
\label{subsec:bh}

In this test we consider different black hole spacetimes. A 
spherical black hole was evolved with puncture and Kerr-Schild 
initial data; evolutions were performed with both the gamma 
driver and asymptotically harmonic shift.

\begin{figure}[t]
  \begin{center}
    \includegraphics[width=0.5\textwidth]{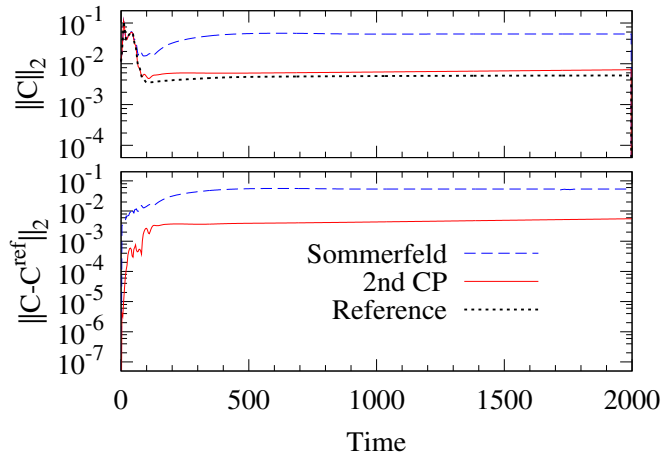}
    \caption{ \label{fig:bh_C} Constraint violation in black 
      hole spacetime test. Upper panel: The 2-norm of the 
      constraint monitor is showed in time for different BCs 
      implemented. The same quantity for the reference 
      simulation is showed (black dotted line).
      Bottom panel: The 2-distance of the constraint monitor 
      with the reference simulation is showed in time for 
      different BCs implemented.}
  \end{center}
\end{figure}

\begin{figure}[t]
  \begin{center}
    \includegraphics[width=0.5\textwidth]{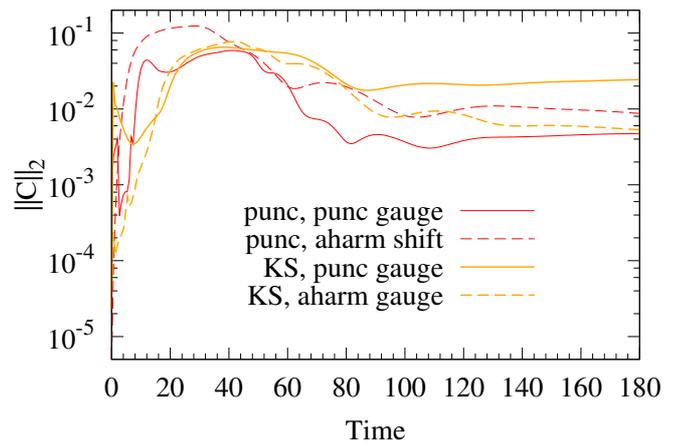}\\
    \caption{ \label{fig:bh_gauge} Constraint violation in black 
      hole spacetime test, comparison of different initial data 
      and gauges. The 2-norm of the constraint monitor is showed
      in time for 2nd order CPBCs. The evolutions refer to puncture 
      initial data showed with puncture shift (red solid line) and 
      asymptotically harmonic shift (red dashed line) and to 
      Kerr-Schild initial data (evolved with excision) with
      puncture shift (thick orange solid line) and 
      asymptotically harmonic shift (thick orange dashed 
      line).}
  \end{center}
\end{figure}

We focus first on the evolution of a single puncture. 
Figure~\ref{fig:bh_C} shows that the behavior of the constraint 
monitor, computed outside the apparent horizon, is analogous 
to what we found in the flat and star spacetime tests. As the 
solution asymptotes to the stationary trumpet 
slice~\cite{Hannam:2006vv} we find more constraint violation 
than in the evolution of the star. The resolution employed in 
the simulations for the figure is quite moderate and the outer 
boundary very close, even compared with a 3D code. Nonetheless 
the CPBCs perform quite well and significantly improve the 
numerical solution with respect Sommerfeld. No big differences 
were found when adopting either the gamma driver shift or the 
asymptotically harmonic one. As shown in Figure~\ref{fig:bh_gauge} 
the puncture shift (red solid line) does even better than the 
asymptotically harmonic (red dashed line) after $t\sim60$. To 
the best knowledge of the authors this is the first time that 
the asymptotically harmonic shift has been used in the evolution 
of puncture data.

To further assess the robustness of our BCs we evolve the
spherical black hole with Kerr-Schild initial data:
\begin{align}
ds^2&=-\left(1-\frac{2M}{r}\right)\,dt^2 + \frac{4M}{r}\,dt\,dr 
+\left(1+\frac{2M}{r}\right)\,dr^2 \nonumber\\
&+ r^2\, d\Omega^2\,.
\end{align}
The excision surface is at $r=1.9M$ and simple 
extrapolation is used at the inner boundary. We stress that this 
test is initially more demanding for the gamma driver shift. Since 
the line-element is not written in a manifestly conformally flat 
form, the shift immediately evolves everywhere in space, which is
not the case with the gamma driver shift and puncture data. 
Figure~\ref{fig:bh_gauge} shows that the results for the constraint 
monitor obtained with our CPBCs are comparable to the puncture case 
with both the puncture shift (orange thick solid line) and the 
asymptotically harmonic (orange thick dashed line). In this case 
the latter perform better. Moreover, at this resolution, the bigger 
spurious reflections produced by Sommerfeld conditions combined 
with the close boundary used ($r^{\rm out}=20$) cause the 
simulations to crash. To achieve stable evolutions with the 
Sommerfeld conditions we find it necessary to use a higher 
resolution and a more distant outer boundary.

\section{Conclusion}
\label{section:Conclusion}

For numerical applications of free-evolution schemes in general 
relativity there is a strong motivation to construct constraint 
preserving boundary conditions. Without such conditions, 
constraint violations may appear at the boundary of the 
numerical domain and swamp the numerical solution. In the worst 
case such errors could be interpreted as real physics. In 
this study we have constructed constraint preserving conditions 
for the Z4c formulation of the Einstein equations with 
variations of the popular puncture gauge. 

We demonstrated well-posedness of the resulting initial 
boundary value problem for the constraint subsystem
on a four dimensional compact manifold in the high-frequency approximation.
Since we are only interested in  the constraint absorption properties of 
our boundary conditions we just analyzed the initial boundary value problem 
of the a spherical reduction of the Z4c system for a 
special choice  of free parameters of the gauge condition. 
One may be able to expand our calculations for the 
asymptotically harmonic shift condition by verifying the 
existence of a suitable symmetrizer in a neighborhood of the 
point $p$ about which we perturb to  reach the high-frequency 
limit. Unfortunately there is not a clear method for extending 
our calculations beyond the high-frequency limit with the 
standard gauge choice. Even in the high-frequency approximation 
we find that the Laplace-Fourier method becomes cumbersome. The 
key problem is that the cascade approach of Kreiss and Winicour 
fails with the puncture gauge.

In order to build a body of evidence for the well-posedness of 
the initial-boundary value problem with the standard puncture 
gauge we therefore performed numerical evolutions of various 
spherical initial data sets. Roughly speaking our approach for 
the implementation of the boundary conditions is to rewrite them 
as closely as possible to Sommerfeld conditions most commonly 
used in BSSN evolutions. We will report our method in detail 
elsewhere. Since the underlying formulation is not symmetric 
hyperbolic we are not able to make a summation-by-parts approach 
to the implementation. Therefore numerical stability is most 
straightforwardly established by studying toy problems 
mathematically and performing existence numerical tests on 
the numerical system. We compared the behavior of the system 
with the standard puncture gauge and the asymptotically harmonic 
gauge. We find very similar features in all tests. We demonstrated 
that with our approach to the numerical implementation we can 
achieve clean pointwise convergence even in reflected constraint 
violation. We also found, in agreement with previous studies, 
that high-order boundary conditions are able to absorb outgoing 
constraint violation much more effectively than first order 
conditions. 

There are two obvious places in which we would like to strengthen 
our results. Firstly it is desirable to extend our well-posedness 
results, especially in the case of the standard puncture gauge. 
However in order to do this a different mathematical approach 
will probably be necessary. Secondly, that the numerical tests 
were performed in spherical symmetry is an obvious drawback which 
we aim to address shortly. Preliminary tests in 3D with the BAM 
code indicate that additional tangential terms in the boundary 
conditions, which we have discarded in this paper, are required 
in order to get a stable evolution.

\acknowledgments

The authors wish to thank Bernd Br\"ugmann, 
Ronny Richter and Olivier Sarbach for helpful discussions and 
comments on the manuscript. The authors thank Luisa Buchman 
for a clarification about the BCs employed in the SpEC code. 
The authors enjoyed several fruitful discussions at the 
``$scri^+$'' fair-trade cafe in Jena, and would like to thank 
the friendly staff there. This work was supported in part by 
DFG grant SFB/Transregio~7 ``Gravitational Wave Astronomy''.

\appendix 

\section{Well-Posed problems}
\label{section:well-posed problems}

The well-posedness of the IBVP is the requirement that 
for given initial and boundary data an unique 
solution should exist and it should depend
continuously on the data~(see {\it e.g.}~\cite{Friedrich99}). 
In this section we present a short
review  of the theory for  first order systems to prove well-posedness
This theory developed by Kreiss~\cite{Kreiss70} gives us necessary 
and sufficient conditions for well-posedness of the IBVP for strictly hyperbolic
systems. The theory was extended to hyperbolic systems of constant
multiplicity by Agranovich~\cite{Agranovich:1972}. The discussion 
presented here follows closely that of 
in~\cite{Kreiss70,Kreiss:2006mi,Rinne:2006vv}.

Consider a strongly hyperbolic system of equations of motion
\begin{equation}
\partial_t {u}= {B}\,\partial_{x^1} u
+\sum_{A=2}^n{C}^A\,\partial_A {u}\,,
\label{eq:def-system}
\end{equation}
with constant coefficients on the half-space $t\ge 0$, 
$x^1 \ge 0$ and $-\infty < x^2, \dots, x^n<\infty$. Here
${u}$ is a $n$-dimensional vector and  ${B}$ and 
${C}^i$ are $n\times n$ constant matrices. By assuming that
${B}$ is non-singular, it can be rewritten as
\begin{equation}
{B}=\left(
\begin{array}{cc}
-{\Lambda}^I&0\\
0&{\Lambda}^{II}\\
\end{array}
\right)\,,
\label{eq:formA}
\end{equation}
with ${\Lambda}^I$ and ${\Lambda}^{II}$ real and
positive definite diagonal matrices of order $m$ and $n-m$ respectively.
We imposed $m$ BCs at $x^1=0$ in the form
\begin{equation}
\left.{L}\,{u}^I(t,x)\right|_{x^1=0}\hateq {g}(t,x)\,,
\label{eq:BC}
\end{equation}
where ${L}$ is a $m\times m$ constant matrix  
and $g={g}(t,x^2,\dots,x^n)$ is a given boundary data vector. 
For simplicity, we consider trivial  initial data $u(0,x)=0$. 
In the following we solve  the IBVP~(\ref{eq:def-system}-\ref{eq:BC}) by
performing a Laplace-Fourier transformation with respect to the directions 
$t$ and $x^A$ tangential to the boundary $x^1=0$. 

Let $\tilde u=\tilde u(s,x^1,\omega^A)$ denote the Laplace-Fourier
transformation of $u(t,x)$. Then, $\tilde u$ satisfies
the ordinary differential system
\begin{eqnarray}
&&\hspace{-1cm}
\partial_x\tilde u=M(s,\omega)\,\tilde u\,,
\qquad\textrm{on}\,x\in(0,\infty)\,,
\label{eq:lapl-four}\\
&&\hspace{-1cm}
 {L}\tilde{ {u}}^I
\hateq\tilde{ {g}}\,,
\hspace{1.9cm}\textrm{at}\,x\hateq0\,,
\label{eq:BC-lapl-four}
\end{eqnarray}
where  $\tilde{ {g}}$ denotes the Laplace-Fourier  transformation  
of $g$ and 
\begin{equation}
M(s,\omega)= B^{-1}\,(s\,\mathbb{I}_{n\times n}+i\,\omega_A\, {C}^A)\,.
\end{equation}
If $\lambda_i$ and $e_i(s,\omega)$ are the corresponding eigenvalues 
and eigenvectors of $M$ then the general $L_2$ solution 
(functions which are quadratically integrable) of (\ref{eq:lapl-four}) 
is given by
\begin{equation}
\tilde{ {u}}=\sum_{i=1}^m\sigma_i\,{e}_i(s,\omega)\,
\,\textrm{exp}(\lambda_i\,x^1)\,,
\label{eq:solution_lap-four}
\end{equation}
where $\sigma_i$ are complex integration 
constants~\footnote{If the matrix $M$
does not have a complete set of eigenvectors then the integration constants
have to be replaced with polynomials in $x^1$.}.
These constants are  determined by the boundary 
conditions.
By substituting~(\ref{eq:solution_lap-four}) into the 
expression~(\ref{eq:BC-lapl-four}) we obtain a system of $m$ linear 
equations for the unknown $\sigma_i$. This system can be written in 
the form
\begin{equation}
\mathbb{D}(s,\omega)\, {\sigma}\hateq\tilde{ {g}}\,,
\label{eq:bc_general}
\end{equation}
where $\mathbb{D}(s,\omega)$ is a $m\times m$ matrix. 
Let us consider for a moment homogeneous  BCs 
$ {\tilde{g}}=0$ and suppose that  there is $s$ with $\re(s)=\eta>0$
such that $\textrm{Det}\,\mathbb{D}=0$.
It means that~(\ref{eq:bc_general}) has a non-trivial solution 
and therefore, the  solution of the IBVP (\ref{eq:def-system}-\ref{eq:BC}) 
is
\begin{equation}
 {u}(t,x^i)=\tilde{ {u}}(x^1)\,
\textrm{exp}\left(s\,t+i\,\omega_A\,x^A\right)\,.
\label{sol:lf}
\end{equation}
This implies, by homogeneity of the system~(\ref{eq:lapl-four}-\ref{eq:BC-lapl-four}), that
\begin{equation}
 {u}(t,x^i)=\tilde{ {u}}(\theta\,x^1)\,
\textrm{exp}\left(\zeta\,s\,t+i\,\zeta\,\omega_A\,x^A\right)\,.
\label{sol:lf2}
\end{equation}
is also a solution for any constant $\zeta>0$. By increasing 
that constant arbitrarily one can find a solution which grows exponentially.
Therefore, the IBVP is not well-posed. 
We conclude that the so-called {\it determinant condition}
\begin{displaymath}
\textrm{Det}\,\mathbb{D}\neq0\,,\qquad\textrm{for}\qquad \eta>0\,, 
\end{displaymath}
is a necessary condition for well-posedness. 

Next, let us consider the inhomogeneous BCs. Since the 
determinant condition  is satisfied we can 
solve (\ref{eq:bc_general}) for the integration constants. What remains
to be shown is that solution~(\ref{eq:solution_lap-four})
can be bounded in terms of the data given at the boundary,
\begin{equation}
|\tilde{ {u}}(s,0,\omega)|
\leq C\,|\tilde{ {g}}(s,\omega)|\,,
\label{eq:boundary_stable}
\end{equation}
where $C>0$ is an independent constant of $s$ and $\omega$. 
Using (\ref{eq:boundary_stable}) and by inverting 
the Laplace-Fourier  transformation it is possible to show that the 
estimate
\begin{equation}
\int_0^T\|u(t,\cdot)\|^2\,dt\leq \delta\int_0^T\left.
\|g(t,\cdot)\|^2\right|_{x^1=0}\,dt\,,
\label{eq:estWoutS}
\end{equation}
holds, {\it i.e.} we can estimate the $L_2$ norm of the solution
in terms of the $L_2$ norm of the given boundary data.
Here the constant $\delta$ is independent of the boundary data.
Systems whose solution~(\ref{eq:estWoutS}) satisfies
this estimate are called {\it boundary stable}~\cite{Gustafsson95,Kreiss70,kreiss2001}.

Kreiss has shown in~\cite{Kreiss70,Kreiss:2006mi} that if the system
(\ref{eq:def-system}) is strictly hyperbolic, the 
condition~(\ref{eq:boundary_stable})
implies that there is  a symmetrizer $ {H}= {H}(s',\omega')$ 
such that
\begin{enumerate}
\item[I.] $ {H}(s',\omega')$  is a smooth bounded function 
that depends of $(s',\omega')$,
\item[II.] there exists a constant $\delta > 0$ such that
\begin{displaymath}
H\,M+M^*\,H
 \geq \delta\,\eta\, \mathbb{I}\,,
\end{displaymath}
for all $\eta > 0$ and all $\omega\in \mathbb{R}$,
\item[III.]
there are constants $\delta_2 > 0$ and $C > 0$ such that
\begin{displaymath}
\left< \tilde u, {H}\,\tilde u \right> \, \geq \delta_2 |\tilde u|^2 
 - C|\tilde g|^2\,,
\end{displaymath}
for all $\tilde u$ satisfying the boundary condition~(\ref{eq:bc_general}),
\end{enumerate}
where $s'=s/\kappa$, $\omega'=\omega/\kappa$ and $\kappa=\sqrt{|s|^2+\omega^2}$
and $\omega^2={\omega_y^2+\omega_z^2}$. We have denoted
the standard scalar product by $\left<,\,\cdot\,,\right>$   and 
$|\cdot|$ its corresponding norm. This symmetrizer allows us obtain an 
estimate of the solution  of the system for which we add
a source term $F(t,x)$ in the right of~(\ref{eq:def-system}). 
In particular, according 
to~\cite{Kreiss:2006mi}, we obtain an estimate of the form
\begin{eqnarray}
&&\hspace{-1cm}
\int_0^T\| {u}(t,\cdot)\|^2\,dt 
+\int_0^T \left\| {u}(t,\cdot)\|^2\right|_{x^1=0}\,dt\leq\nonumber\\
&&\hspace{-1cm}
\delta\,\left(\int_0^T\| {F}(t,\cdot)\|^2\,
dt\, +
 \int_0^T\left.\| {g}(t,\cdot)\|^2\right|_{x^1=0}\,dt\right)\,,
\label{eq:gen_estimate}
\end{eqnarray}
where the constant $\delta>0$ 
is independent of the boundary data $ {g}$ or the source term 
${F}$. 

\section{Toy Model }       
\label{section:Toy_model0} 
Consider the wave equation 
\begin{eqnarray}
\left[\partial^2_0 -\mu^2\,\partial^l\partial_l\right]\,
U\left(t,x^i\right)\,=\,F\left(t,x^i\right)\,,
\label{eq:eqMU}
\end{eqnarray}
on the half-space $t\ge 0$, $x \ge 0$ and $y$ and 
$z\in(-\infty,\infty)$ with trivial initial data.
We impose  BCs 
at  $x\,\hateq\,0$ of the form
\begin{eqnarray}
\left[\partial_0-\mu\,\partial_x\right]\,U(t,x^i)
\hateq h\,,
\label{eq:bc_phi}
\end{eqnarray}
where $h$ are given boundary data and $\p_0$ is the time derivative 
along the coordinate time in the linear regime.

According to~\cite{Ruiz:2007hg}, let us denote to $\tilde U$  as the
Laplace-Fourier  transformation  of 
$U(t,x^i)$ with respect to the directions $t$, $y$ and $z$ 
tangential to the boundary then in the background~\eqref{eq:backg-metricP}, 
$\tilde U$ satisfies
\begin{eqnarray}
&&\hspace{-0.8cm}
\left[\left(\mu^2-\mathring\beta^2\right)\p_x^2
+ 2\,\mathring\beta\,s\,\p_x
-(s^2+\mu^2\,\omega^2)\right]\tilde U=\tilde F\,,
\label{eq:LF-u}\\
&&\hspace{-0.8cm}
\left[s-\left(\mu+\mathring{\beta}
\right)\,\p_x\right] 
\tilde U\hateq \tilde h\,,
\label{eq:BC-u}
\end{eqnarray}
where $\tilde{F}$  and $\tilde{h}$ denote the Laplace-Fourier  transformations 
of $F$ and $h$ respectively. In order to apply the theory
presented in the appendix~\ref{section:well-posed problems}, one can
rewrite the above system as a first order one by introducing  the variable~\cite{Ruiz:2007hg,Kreiss:2006mi} 
\begin{eqnarray}
D\tilde U= \frac{1}{\kappa}\,\left(\p_x\tilde U+
\gamma^2_\mu\,\mathring\beta\,s\,\tilde U\right)\,,
\end{eqnarray}
where $\gamma_\mu=1/\sqrt{\mu^2-\mathring\beta^2}$. Therefore,
the system~(\ref{eq:LF-u}) can be rewritten
in the form
\begin{eqnarray}
&&\partial_x\tilde{W} = M(s,\omega)\, \tilde{W} 
+ \tilde{f}\,,
\label{eq:generalform-U}\\
&&L(s,\omega)\tilde W=\tilde g\,,
\label{eq:generalform-U-BC}
\end{eqnarray}
where we have defined
\begin{displaymath}
\tilde{W} = \left( \begin{array}{c} \tilde U
\\ D\tilde{U} \end{array} \right)\,,
\qquad
\tilde{f} = \frac{\gamma_\mu^2}{\kappa} 
 \left( \begin{array}{c} 0 \\ \tilde{F} \end{array} \right)\,,
\end{displaymath}
and
\begin{eqnarray}
M(s,\omega) &=& \kappa\,\left( \begin{array}{cc}
-\gamma_\mu^2\,\mathring\beta\,s'  & 1 \\
\mu^2\,\gamma^4_\mu\,\lambda^2
 & -\gamma_\mu^2\,\mathring\beta\,s'
\end{array} \right)\,,\\
L(s,\omega)&=&(\mu\,s',-\gamma^{-2}_\mu)\,,
\label{eq:LF-Mat}
\end{eqnarray}
with $\lambda^2=s'^2 +\gamma^{-2}_\mu\, \omega'^2$. 
The  $L_2$ solution of the homogeneous system
(\ref{eq:generalform-U}) is given by
\begin{equation}
\tilde{W}(s,x,\omega) = \sigma\,e^{(\tau^- x)}\,e^-\,.
\label{eq:general_sol_phi}
\end{equation}
where $\tau^-$ is the eigenvalue of $M$ with $\re(\tau^-)<0$
and $e^-$ its corresponding eigenvector.
Introducing~(\ref{eq:general_sol_phi}) into the boundary 
we have
\begin{equation}
\mu\,\left(\, s' + \sqrt{s'^2 + \gamma^{-2}_\mu\,\omega'^2}\, 
\right)\sigma = \tilde g\,.
\label{eq:bc_est-phi}
\end{equation}
According to~\cite{Kreiss:2006mi,Ruiz:2007hg},
one can show that there is a constant $\delta>0$ such that
\begin{equation}
\left| s' + \sqrt{s'^2 +\gamma^{-2}_\mu\, \omega'^2}\, 
\right|\geq \delta\,.
\label{eq:bc_est-bc}
\end{equation}
Provided that the eigenvector $e^-$ in the 
solution~(\ref{eq:general_sol_phi}) is normalized in a way that it 
remains finite as $\omega$ or $|s|$ goes to zero or infinity, there 
is a constant $C> 0$ such that
\begin{equation}
|\tilde{W}(s,0,\omega,)|\leq\,C\,|\tilde{g}|\,,
\label{eq:Bon-stab-phi}
\end{equation} 
for all $s\in \mathbb{C}$ with $\eta>0$, and $\omega\in \mathbb{R}$.
Therefore, the  system is boundary stable. So, we should consider the 
existence of a symmetrizer $H=H(s',\omega')$ and use  it to get an 
energy estimate for the full problem. According to~\cite{Ruiz:2007hg}, 
it can be shown that the following estimate
\begin{eqnarray}
\eta\,\int\limits_0^\infty 
 \left( | \kappa\,\tilde U|^2 
+ |\partial_x\tilde{U}|^2 \right) \,dx
+ \left. \left( | \kappa\,\tilde{U}|^2 
+ |\partial_x\tilde{U}|^2 \right) \right|_{x=0}
&&\nonumber\\&&
\hspace{-6.cm}
\leq C'\,\left[\frac{1}{\eta}\,\int\limits_0^\infty 
\left|{\tilde F}\right|^2 \,dx 
+ \left|\tilde{h}\right|^2 \right]\,,
\label{eq:phi-estimate}
\end{eqnarray}
holds for some constant $C' > 0$. Therefore, by inverting the 
Laplace-Fourier transformation and using  the Parseval's relation, one 
obtain the  estimate~(\ref{eq:gen_estimate}) for the solution in terms 
of the  $L_2$ norm of the boundary data.

One can generalize the boundary condition~(\ref{eq:bc_phi}) to higher order
BCs. It has been shown that such conditions reduce the 
amount of reflections at the boundary~\cite{Buchman:2007pj,Rinne:2008vn}. 
Thus, we impose high order BCs  of the form
\begin{eqnarray}
\left[\partial_0-\mu\,\partial_x\right]^{m+1}\,U(t,x^i)
\hateq h\,,
\label{eq:bc_phi-ho}
\end{eqnarray}
with $m\ge 1$. Following~\cite{Ruiz:2007hg} it is  possible rewrite the 
previous conditions  as
\begin{equation}
\mathcal{L}^{m+1}\tilde U\hateq
\left[\frac{\mu-\mathring\beta}{\kappa}\right]^{L+1}\,h\,,
\label{eq:high-orderwave}
\end{equation}
where the linear operator $\mathcal{L}$ is defined by 
$\mathcal{L}=(\mu-\mathring\beta)\,s'-\p_x/{\kappa\,\gamma_\mu^{2}}$.
Using the equation of motion~(\ref{eq:generalform-U}), we can rewrite the above 
condition in algebraic form. Note that
\begin{equation}
\mathcal{L}\left(
\begin{array}{c}
\tilde U\\D\tilde U
\end{array}
\right)=
A\,\left(
\begin{array}{c}
\tilde U\\D\tilde U
\end{array}
\right)
-\frac{1}{\kappa^2}\,\left(\begin{array}{c}
0\\\tilde F
\end{array}
\right)\,,
\end{equation}
where the matrix $A$ is given by
\begin{eqnarray}
A&=& \left( \begin{array}{cc}
\mu\,\,s'  & -\gamma_\mu^{-2} \\
-\mu^2\,\gamma^2_\mu\,\lambda^2
 & \mu\,s'
\end{array} \right)\,.
\label{eq:A-BC}
\end{eqnarray}
It has been shown in~\cite{Ruiz:2007hg} that by iteration
the boundary condition~(\ref{eq:high-orderwave}) can be rewritten in the form
\begin{equation}
L(s,\omega)=\frac{1}{2}\,\left(a_+^{m+1}+a_-^{m+1}, -
\,\frac{a_+^{m+1}-a_-^{m+1}}{\mu\,\lambda\,\gamma_\mu^{2}}\right)\,,
\label{eq:h-bc}
\end{equation}
where $a_{\pm}=\mu\,(s'\pm\lambda)$ are the eigenvalues of $A$. 
The  $L_2$ solution of the homogeneous wave equation
(\ref{eq:generalform-U}) is given by~(\ref{eq:general_sol_phi}).
Nevertheless, the integration constant $\sigma$ satisfies 
$a_+^{m+1}\,\sigma = \tilde g$. It can be shown that the
system~(\ref{eq:generalform-U})  with BCs~(\ref{eq:h-bc})
is boundary stable and, according to~\cite{Ruiz:2007hg}, the solution
satisfies the following estimate
\begin{eqnarray}
&&\hspace{-0.2cm}
\eta\,\int\limits_0^\infty\, 
\sum\limits_{j=0}^{m+1} \left|\kappa^{(m+1)-j}\,\p_x^j
\tilde U\right|^2\,dx+ \left. 
\sum\limits_{j=0}^{m+1} \left|\kappa^{(m+1)-j}\,\p_x^j\tilde U
\right|^2\right|_{x=0}
\nonumber\\
&&\hspace{1.5cm}\leq C\left[ 
\frac{1}{\eta} \,\int\limits_0^\infty
\sum\limits_{j=0}^{m-1} \left| \kappa^{m-j}\partial_x^j
\tilde F\right|^2\, dx\right.
\nonumber\\
&&\hspace{1.5cm}  + \left.\left. \sum\limits_{j=0}^{m-1}
\left| \kappa^{(m-1)-j}\p_x^j\tilde F\right|^2 
\right|_{x=0} + \left|\tilde{h}\right|^2 \right]\,,
\label{eq:full-estimate-W}
\end{eqnarray}
for some strictly positive constant ${C} > 0$. Thus, by inverting 
the Laplace-Fourier transformation we can estimate 
the $L_2$ norm of  higher derivatives of the solution in 
terms of given data.


\section{Implementation of boundary conditions in spherical 
symmetry}
\label{App:BC_imp}


In this appendix we describe the numerical implementation of
the second order constraint preserving boundary 
conditions~(\ref{eq:general_CPBCs}-\ref{eq:general_BCs_gauge_first}) 
in spherical symmetry.

We write the spherical line-element as 
\begin{align}
\textrm{d}s^2=\chi^{-1}\tilde{\gamma}_{rr}\textrm{d}r^2
+\chi^{-1}\tilde{\gamma}_Tr^2\textrm{d}\Omega^2,
\end{align}
where $\textrm{d}\Omega^2=\textrm{d}\theta^2
+\sin^2\theta\textrm{d}\phi^2$. Similarly we evolve 
$(\hat{K},\tilde{A}_{rr},\tilde{A}_{T})$ for the extrinsic curvature. 
In spherical symmetry the algebraic 
constraints~\eqref{eq:Conf_Constr_2} are
\begin{align}
D=\log \left(\tilde{\gamma}_{rr}\tilde{\gamma}_{T}^2\right)=0,
&\quad 
T = \frac{\tilde{A}_{rr}}{\tilde{\gamma}_{rr}}
+2 \frac{\tilde{A}_{T}}{\tilde{\gamma}_{T}}=0.
\end{align}
Using the linearized equations of motion for the system, we 
rewrite the the boundary conditions as 
\begin{align}
\p_t\Theta&\hateq-\p_r\Theta-\frac{1}{r}\Theta,
\label{eq:sph_bc_1st}\\
\p_t\tilde{\Gamma}^r&\hateq-\frac{2}{\sqrt{3}}\p_r\tilde{\Gamma}^r
-\frac{2}{\sqrt{3}r}\tilde{\Gamma}^r
-\frac{4}{3r^2}\beta^r\nonumber\\
&-\frac{1}{3}\p_r\Theta-\frac{2}{3}\p_r\hat{K}
,\\
\p_t\hat{K}&\hateq -\sqrt{2}\,\p_r\hat{K} 
-\frac{\sqrt{2}}{r}\,\hat{K} 
+ \frac{1}{r}\,\p_r\alpha,\\
\tilde{A}_{rr}&\hateq-2\p_r\tilde{A}_{rr}-\frac{6}{r}\tilde{A}_{rr}
-\p_r\p_r\tilde{\gamma}_T-\frac{1}{2}\p_r\tilde{\Gamma}^r\nonumber\\
&-\frac{2}{3}(-2+\sqrt{2})\p_r\hat{K}+\frac{1}{3}\p_r\Theta
-\frac{2}{r^2}(\tilde{\gamma}_{rr}-\tilde{\gamma}_{T})\nonumber\\
&+\frac{1}{r}\big(\frac{5}{2}\p_r\tilde{\gamma}_{rr}+2\p_r\alpha
-\frac{1}{3}\Theta-\frac{2\sqrt{2}}{3}\hat{K}-2\tilde{\Gamma}^r
\nonumber\\
&-\p_r\tilde{\gamma}_T-\p_r\chi\big),\label{eq:sph_bc_last}
\end{align}
where here, for brevity, we have linearized around flat space. 
The boundary condition for $\tilde{A}_T$ can be obtained by using 
the algebraic constraints. Note the similarity with the Sommerfeld 
boundary condition. For the numerical implementation we populate 
ghostzones for each gridfunction $f_i$ by sixth order extrapolation 
\begin{align}
f_{N+i}&=6f_{N+i-1}-15f_{N+i-2}+20f_{N+i-3}-15f_{N+i-4}\nonumber\\
&+6f_{N+i-5}-f_{N+i-6},
\end{align}
in order to approximate derivatives and compute Kreiss-Oliger 
artificial dissipation at the boundary. Here $N$ denotes the boundary
point. For the variables $(\Theta,\tilde{\Gamma}^r,\hat{K},
\tilde{A}_{rr},\tilde{A}_{T})$ we simply replace the standard 
evolution equations with~(\ref{eq:sph_bc_1st}-\ref{eq:sph_bc_last}) 
at the boundary. The remaining variables are evolved according to 
their standard equation of motion at the boundary.

\bibliographystyle{apsrev}           
\bibliography{refs/references}{}     

\end{document}